# Mid-Infrared Cross-Comb Spectroscopy

Mingchen Liu[1], Robert M. Gray[1], Luis Costa[1], Charles R. Markus[2], Arkadev Roy[1], Alireza Marandi[1, *]

[1]Department of Electrical Engineering, California Institute of Technology, Pasadena, California, 91125, USA
[2]Division of Chemistry and Chemical Engineering, California Institute of Technology, Pasadena, California, 91125, USA
[*]marandi@caltech.edu

**Dual-comb spectroscopy has been proven a powerful tool in molecular characterization, which remains challenging to implement in the mid-infrared region due to difficulties in the realization of two mutually locked comb sources and efficient photodetection. Moreover, the detection capability of dual-comb spectroscopy is fundamentally limited by the strong excitation background and detector saturation. Here we introduce a variant of dual-comb spectroscopy called cross-comb spectroscopy, in which a mid-infrared comb is upconverted via sum-frequency generation with a near-infrared comb of a shifted repetition rate and then interfered with a spectral extension of the near-infrared comb. We show that cross-comb spectroscopy can have superior signal-to-noise ratio, sensitivity, dynamic range, and detection efficiency compared to other dual-comb-based methods and avoid the limits of the background excitation and detector saturation. We experimentally demonstrate a proof-of-concept measurement of atmospheric $CO_2$ around 4.25 μm, with a 233-cm$^{-1}$ instantaneous bandwidth, 28000 comb lines, a single-shot SNR of 167 and a figure of merit of $2.4×10^6$ Hz$^{1/2}$. Cross-comb spectroscopy can be realized using up- or down-conversion and offers an adaptable and powerful spectroscopic method outside the well-developed near-IR region. This approach opens new avenues to high-performance molecular sensing with wavelength flexibility, which can impact a wide swath of applications.**

Dual-comb spectroscopy (DCS), based on two mutually locked frequency comb (FC) sources in the same wavelength range, has become a compelling alternative to traditional Fourier-transform infrared spectroscopy (FTIR) with advantages in resolution, precision, sensitivity, speed, and bandwidth[1–3]. Over the past decade, significant efforts have focused on its extension to the mid-infrared (MIR) spectral region (3-25 μm)[4–9], where strong molecular signatures are located, making it promising for numerous applications in physical, chemical, biological, and medical sciences or technologies. However, generating two mutually locked broadband frequency comb sources in the MIR has posed a significant challenge. In addition, photodetectors in the MIR usually suffer from lower sensitivity, higher noise, and slower response times, and generally require cooling, compared to their well-developed near-infrared (NIR) counterparts. Moreover, photodetection above 13 μm[10] remains a significant challenge.

On the other hand, the signal-to-noise ratio (SNR), sensitivity (detection limit) and dynamic range (DR) of DCS have been limited by noise from strong excitation background[11–13], especially for detection of trace molecules. To detect a weaker absorption, one can always apply a higher excitation power, but the stronger background signal from it will in turn decreases the dynamic range, which is an undesirable inherent tradeoff of DCS. Additionally, the sensitivity is still fundamentally limited by the detector saturation. Therefore, although significant progress has been made toward broadband and high-power (>100 mW) MIR frequency combs[14–20], the MIR DCS cannot take full advantage of such sources since typical MIR detectors saturate at ~ 1 mW.

To overcome those obstacles, one effective path is to upconvert the MIR FC to the NIR region using short pulses and capture the wealth of molecular information available in the MIR with NIR photodetectors. Electro-optic sampling (EOS) is one recent successful example of this approach[13,21], in which ultrashort NIR pulses are used to directly detect the electric field of MIR pulses in the time domain. However, this method necessitates extremely short NIR pulses with durations shorter than the optical cycle of the carrier frequency of the MIR pulses[22,23], whose generation and dispersion control require substantial efforts. Besides, the detection is based on field-dependent polarization rotation of the NIR sampling pulses, which adds extra complexity to the system. Moreover, although temporal gating by ultrashort NIR pulses has been demonstrated to increase the sensitivity and dynamic range of the measurement since it can isolate the weak free induction decay (FID) signal from the strong MIR excitation background[13], the detectors used in EOS are still under a background signal from the strong NIR pulses themselves, the saturation of which can still limit the detection.

In addition to EOS, one can also upconvert the MIR frequency comb using a high-power NIR continuous-wave (C.W.) laser and perform standard DCS in the NIR region[24]. Nonetheless, this method has not yet been demonstrated to exhibit a favorable signal-to-noise ratio (SNR) and bandwidth compared to direct MIR DCS, mainly owing to its low upconversion efficiency because of the low peak power of C.W. lasers. More essentially, this method is still constrained by the above-mentioned limits of the general DCS as there is no temporal gating.

In this work, we introduce a new method named cross-comb spectroscopy (CCS), which can be considered a general form of frequency-converted DCS. Specifically, we demonstrate that the scheme of short-pulse CCS can have better SNR, sensitivity, DR, and detection efficiency compared to DCS or C.W. upconversion DCS, and less experimental complexity compared to EOS. Furthermore, as the detection process has been divided into two parts, upconversion (by local FC) and interference (by readout FC), which can be tuned independently, the detector saturation can be avoided and will no longer limit the detection. In addition, upon a proper setting of power, the detection can fully utilize the DR of the detector, which does not have to be sacrificed for higher sensitivity.

The general idea of CCS is illustrated in Fig. 1a. The spectral information contained in the MIR (target) FC is upconverted to the NIR region via sum-frequency-generation (SFG) with a NIR (local) FC of a slightly shifted repetition rate (shifted by $\delta$). The SFG output is then interfered with the spectral extension of the local FC (readout FC) to transfer the MIR information into the radio frequency (RF) domain. Like DCS, it is possible to map the MIR FC teeth to the RF comb teeth, which are easily accessed with a single NIR detector and RF measurement. To obtain a tooth-resolved absorption spectrum of the target FC, the minimum required aggregate bandwidth of the local and readout FC is equal to the bandwidth of the target FC, which eliminates the need found in EOS for super-short NIR pulse generation and measurement of polarization rotation. We show that CCS can fundamentally have better detection metrics than other demonstrated dual-comb-based techniques, and experimentally demonstrate a measurement around 4 μm with a high temporal SNR and figure of merit (FOM)[12]. Provided suitable comb sources and strong upconversion capability are available, this method can be extended to any wavelength range and promises a superior performance that can break the limits of conventional methods, especially for longer wavelengths.

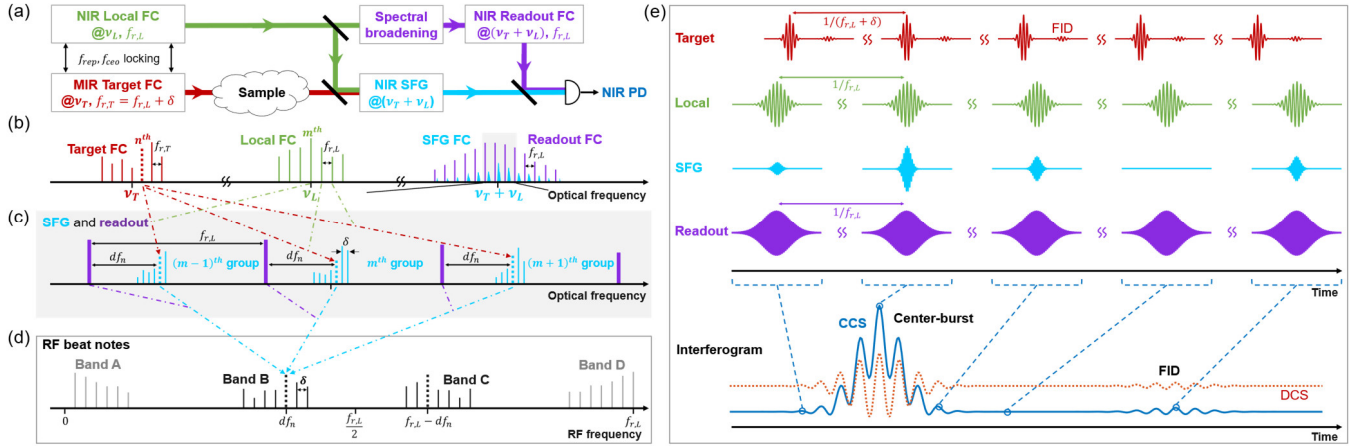

**Fig. 1 Cross-comb spectroscopy. a** Schematic of the setup. $\nu_L$ and $\nu_T$, center optical frequencies of the NIR local FC and MIR target FCs. $f_{ceo}$ and $f_{rep}$, carrier–envelope offset frequency and repetition rate of a FC. $f_{r,L}$ and $f_{r,T}$, repetition rate of the local FC and the target FC. $\delta$, difference between $f_{r,L}$ and $f_{r,T}$. PD, photodetector. **b-d** Principle of the tooth mapping in frequency domain. An example target tooth, together with its corresponding SFG teeth and RF teeth, is denoted by a dashed line to demonstrate the one-to-one mapping. n (m): index of the example tooth of the target FC (local FC). **b** Optical spectra. **c** a zoomed-in view of the grey-shadowed area in (b). Mixed with different local teeth, one target tooth will generate multiple SFG teeth that are distributed in different SFG groups, separated by $f_{r,L}$. Each SFG tooth is the product of a target tooth and a local tooth. The mapping of the $n_{th}$ target tooth is denoted by arrows, as an example. For SFG teeth that are generated by the same target tooth, they share the same distance to their respective

closest readout tooth. For example, $df_n$ denotes the distance for SFG teeth (dashed lines) generated by the $n_{th}$ target tooth. **d** Heterodyne beat notes in the RF domain, obtained by square-law detection of the interference between the SFG FC and readout FC with a single NIR detector. Band B (C) is the result of the beating between SFG teeth with their nearest (second nearest) readout tooth, while band A (D) is the result of the beating between two SFG teeth from the same SFG group (two adjacent SFG groups). The dashed tooth in band B is the unique mapping of the $n_{th}$ target tooth, as is its mirror image in band C. The arrows denote the optical tooth pairs in (c) that contribute to the dashed RF tooth. **e** CCS in time domain. In addition to the typical CCS interferogram (solid blue curve), a typical DCS interferogram (dashed red curve) is plotted for comparison. Note that this illustration describes the case where target, local and readout RF are all short pulses, which is not necessary for general CCS.

## Results

**One-to-one comb tooth mapping.** Figure 1 illustrates the operation principle of CCS using sum-frequency sampling. Each pair of comb teeth from the local FC (green) and target FC (red) generates an SFG tooth at a unique frequency, the set of which is referred to as the SFG FC (blue). The teeth of the SFG FC cluster into different frequency groups that are evenly spaced by the repetition rate of the local FC ($f_{rL}$)[25] and follow specific patterns (Fig. 1c, Supplementary Section 2.2). Within a frequency group, the SFG teeth are separated by δ, and each tooth represents a unique tooth of the target comb. Across all the SFG comb groups, teeth generated by the same tooth of the target comb are all at the same relative frequency position. These characteristics enable one-to-one mapping from the MIR domain to the RF domain. To realize this, a readout FC (purple), which is effectively a spectral extension of the local FC, is employed to beat with the SFG FC on a NIR photodetector. The resultant RF FC contains four distinguishable bands (Fig. 1d)[22]. While band A and D correspond to the envelope of the SFG pulses (intensity cross-correlation between the target FC and local FC) and thus lack spectral information, band B (or its mirror image band C) is a one-to-one mapping from the target FC (multiplied by the local and readout FCs) to the RF FC. To interrogate the spectral response of the sample in the target FC path, one can compare the measured RF band B (or C) with the corresponding sample-free result. Balanced detection can eliminate band A and D since they are common-mode signals, which would double the bandwidth for band B and C. Fig. 1e presents the same principle of CCS in the time domain. The target pulses are sampled by the local pulses, generating SFG pulses, which then interfere with the readout pulses, leading to the RF interferogram. More detailed descriptions can be found in Supplementary Sections 2 and 3.

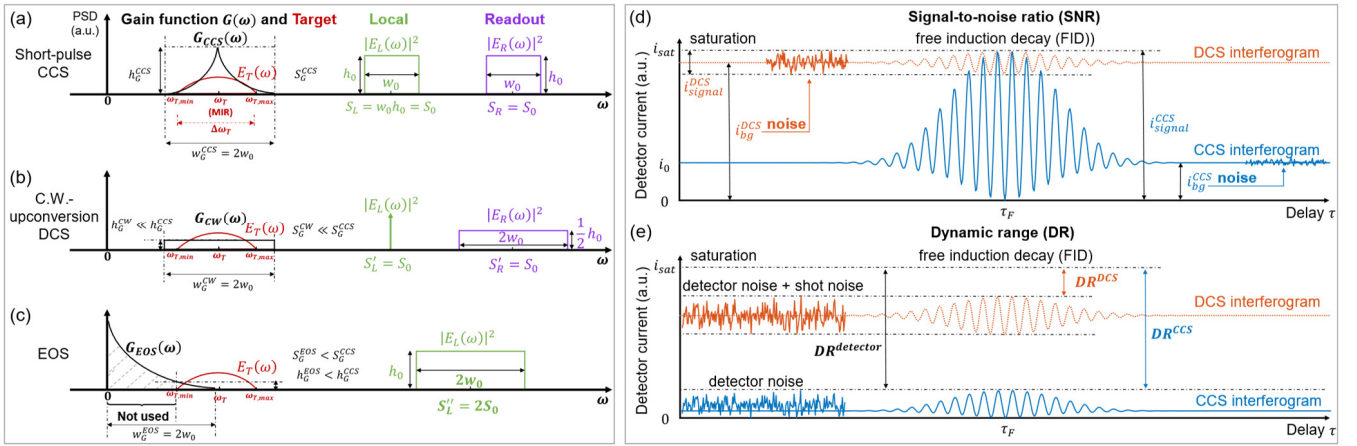

**Fig. 2 Comparison of detection efficiency, bandwidth, SNR and DR between short-pulse CCS and other dual-comb-based techniques. a-c** power gain function $G(\omega)$ for quantification of detection efficiency and bandwidth for three upconversion methods: short-pulse CCS (a), C.W. upconversion DCS (b), and EOS (c). The spectral amplitude of target ($E_T(\omega)$), and spectral intensity of local ($|E_L(\omega)|^2$) and readout FC ($|E_R(\omega)|^2$) are denoted by curves in red, green, and purple, respectively. The instrument response function $H(\omega)$ is the convolution of $E_L^*(-\omega)$ and $E_R(\omega)$, and its spectral intensity ($G(\omega) = |H(\omega)|^2$) is denoted by the black curve for each method. **a** $h_0$, $w_0$ and $S_{L(R)}$ denote the height (PSD), width (bandwidth), and area (total average power) of the local (readout) spectrum. $\Delta\omega_T$, $\omega_{T,min}$, $\omega_T$, $\omega_{T,max}$ denote the bandwidth and minimal, center, and maximum optical frequency of the target spectrum to be detected, respectively. $w_G$, $h_G$ and $S_G$ denote the width, maxima, and total area under the curve of the $G(\omega)$, respectively. **b** In the C.W.

upconversion DCS, we keep the average power of both readout and local to be same as CCS. The "local FC", a C.W. laser, can thus be approximated by a Dirac-delta function since it has a "zero" bandwidth and "infinite" spectral density. To also keep the total optical bandwidth of the local and readout FCs here equal to (a), we double the bandwidth of the readout FC and halve its spectral intensity. **c** In EOS, the functions of the readout and local FCs are accomplished by only one "local FC." Again, to keep the total power and bandwidth same as CCS for a fair comparison, its bandwidth is doubled with an unchanged height. The grey-dashed area of the $G_{EOS}(\omega)$ denotes the part of it not effectively used in detection since it does not overlap with the target spectrum. **d-e** comparison between DCS (red curves) and CCS (blue curves) interferograms at FID. **d** SNR comparison. In DCS, the weak FID must be accompanied by the strong excitation background from the center-burst, which contributes only to noise here, not signal, and thus limits the SNR. Contrarily, CCS is free from such a background, and its signal can form an interference pattern of high visibility, with only a small background noise. $i_{signal}$, the range of the beating signal. $i_{bg}$, the background current; $i_{sat}$, the saturation level of the detector. **e** DR comparison. In DCS, a large part of the detector DR is occupied by the background. The higher the sensitivity (lower detection limit) reached, the larger the background, and the smaller the residual DR of the detector. However, CCS does not have such problem and can make use of the whole detector DR in principle.

**Detection efficiency and bandwidth.** Figure 2 (a)-(c) compares CCS with the other two upconversion methods in terms of detection efficiency and bandwidth. To quantify their performance, we define a power gain function $G(\omega) = |H(\omega)|^2$, where $H(\omega) = E_L^*(-\omega) \otimes E_R(\omega)$ denotes the instrument response function, in which $E_L(\omega)$ and $E_R(\omega)$ denote spectral envelopes (Fourier transforms of a single pulse) of the local and readout FCs, respectively. The width $w_G$, maximum value $h_G$ and total area $S_G$ of $G(\omega)$ can quantify the bandwidth, highest gain, and total gain of the detection, respectively. For short-pulse CCS (panel (a)), we set the spectrum of local and readout FC to be both rectangular functions with the same height, width, and area. Their corresponding $G_{CCS}(\omega)$ (black curve) has a width equal to the sum of their bandwidths, which is required to be larger that of the target spectrum $E_T(\omega)$ (red curve), i.e., $\Delta\omega_T$, to detect the whole target band. As for the C.W. upconversion DCS (panel (b)), the local spectrum shrinks to be like a Dirac-delta function. For a fair comparison, the total average power and optical bandwidth of the local and readout FCs are all kept the same as for CCS. Although $G_{CW}(\omega)$ has the same bandwidth, $h_G^{CW}$ and $S_G^{CW}$ are both much smaller than those of CCS by a factor proportional to the peak power ratio between a short pulse and C.W. laser of the same average power. This detection gain advantage of CCS is rooted in the much higher peak power that (femto-second) short-pulses can offer for the nonlinear wavelength-conversion process, compared to C.W. lasers.

EOS (panel (c)) is slightly different as it uses only one local FC to play the role of both the local and readout FCs of the prior methods (see Fig. 4 and Supplementary section 3). Thus, its $H(\omega)$ is the autoconvolution of $E_L(\omega)$. Again, the total bandwidth and power are kept the same for a fair comparison. Although the nominal bandwidth of $G_{EOS}(\omega)$ is the same as the others, most of it is not effectively utilized (grey-dashed shadow) as it does not overlap with target spectrum ($E_T(\omega)$). To cover the whole $E_T(\omega)$, the local bandwidth is required to be larger than $\omega_{T,max}$ instead of $\Delta\omega_T$, which causes experimental challenges and is an ineffective use of bandwidth. For example, to detect a 30-THz MIR band from 60 THz to 90 THz (5-3.33 μm), EOS would require extremely short local pulses with a bandwidth of at least 90 THz, i.e., 155-245 THz (1.22-1.93 μm) if it is centered around 200 THz (1.5 μm). Meanwhile, CCS only needs a total bandwidth of 30 THz. Moreover, since only the part of $G_{EOS}(\omega)$ that overlaps with $E_T(w)$ contributes to the detection, its effective $h_G$ and $S_G$ are much smaller than their nominal values and depend on the values of $\omega_{T,min}$ and $\omega_{T,max}$. Using the parameters from the previous example, $h_G$ and $S_G$ of EOS are calculated to be 4/9 and 4/27 those of CCS, respectively, even though EOS can be more experimentally demanding. Additional details regarding this section can be found in Supplementary Sections 3 and 4.1.

**Wavelength conversion and temporal gating.** The optical nonlinearity can provide CCS with advantages over DCS in two ways. The first is wavelength conversion. Generally, photodetection in NIR has a better performance, a lower price, and does not require cooling. Also, typical commercial MIR detectors have a cut-off wavelength around 13 μm and are limited for detection of longer wavelengths. Converting MIR information to NIR can get around these limitations.

Second, and more importantly, strong nonlinearity from short pulses gives rise to a temporal gating[13] effect that can endow EOS with better SNR, sensitivity, and dynamic range compared to DCS. To demonstrate such advantages, we compare EOS and DCS directly at the FID part of their interferograms. FID is the sample's reradiation which contains molecular signatures, and isolating it from the center-burst has been demonstrated to provide the absorption spectrum with a better detection performance[13]. Therefore, the SNR of the two techniques at the FID can be a good indicator of their sensing capability for weak absorption signals, the comparison of which is illustrated in Fig. 2d. In DCS, the weak FID beating must be on top of a strong background, most of which comes from the center-burst power which only contributes to the noise here but not the signal. On the contrary, the FID beating in CCS is independent of any extra background, because the FID part of target pulse is temporally isolated from the strong center-burst by the short local pulses (see Fig. 1e). In other words, it is possible to get an "ideal" interference pattern with full interferometric visibility (SNR) at the FID in CCS, while that of the DCS is greatly limited by the center-burst background; the lower the absorption to detect, the higher the excitation power (background) that must be used, and the larger this difference can be. Note that saturation at the center-burst part of the interferogram is ignored for both cases.

The SNR comparison naturally leads to a comparison of sensitivity, i.e., the minimal detectable absorption. In DCS, higher optical power gives higher SNR and sensitivity, which are ultimately limited by the detection saturation (or relative intensity noise (RIN) of the sources)[12]. However, as CCS detects the FID in a "background-free" manner, its fundamental limitation is the nonlinear upconversion capability. That is, the upconverted weak FID signal just has to be stronger than the detector noise (NEP) to be detectable, instead of shot noise or RIN from the strong background in DCS. Therefore, CCS is no longer limited by the detector saturation or extra noise from the strong excitation background, which sets a hard boundary for DCS.

Fig. 2e illustrates the dynamic range comparison. In CCS, as mentioned before, the detector noise limits the weak side of the absorption signal, and the FID detection can utilize the full DR of the detector. Meanwhile, the background in DCS occupies a large part of the detector DR, which inevitably decreases the room for the FID. In fact, a weaker absorption would require a higher excitation (background) to provide sufficient SNR for detection, which occupies more of the detector DR, further decreasing the DR of the FID (absorption). However, this "sensitivity-DR tradeoff" does not exist in CCS as the excitation background is excluded in the FID detection. In principle, CCS can take advantage of the full DR of the detector.

Among the three upconversion methods, C.W. upconversion DCS can only have the advantage on the aspect of wavelength conversion, since the C.W. laser cannot provide any temporal gating and it is still basically a DCS detection. As for EOS, one may think it can have all the advantages discussed above since it uses even shorter pulses for upconversion than CCS. However, this is not fully true because its "local and readout pulses" are actually from the same pulse; therefore, their power cannot be tuned independently. This means that for a very weak FID where very strong pulses are required to upconvert it, the "readout part" of the pulse may already saturate the detector. This will not happen in CCS as the local pulses and readout pulses are separate, meaning that their powers can be tuned independently according to the absorption to detect, and hence the detector saturation can be avoided. In short, the flexibility of short-pulse CCS can utilize the full advantages offered by the optical nonlinearity. A detailed discussion for this section can be found in Supplementary Sections 3 and 4.2.

**Setup.** To experimentally demonstrate CCS in the MIR, we conduct a measurement of atmospheric carbon dioxide ($CO_2$) around 4.25 μm (2349 cm$^{-1}$, antisymmetric stretching mode $\nu_3$). The target FC consists of 50-fs pulses centered at 4.2 μm with 500 mW of average power provided by two-stage cascaded efficient half-harmonic optical parametric oscillators (OPOs), which are intrinsically phase locked to the pump frequency comb (a mode-locked Yb-fiber laser) at 1 μm[14]. The local FC is a NIR FC centered at 1560 nm (a mode-locked Er-fiber laser) with a 100-nm (400-cm$^{-1}$) FWHM bandwidth, 100-fs pulse width and a 200-mW average power (Menlo Systems FC1500-250-WG). The $f_{rep}$ of the target and local FC are both around 250-MHz and are locked to an RF rubidium (Rb) clock with a shift of 1 KHz. The $f_{ceo}$ of the Yb-fiber laser and the Er-fiber laser are both locked via standard f-to-2f techniques. The readout FC is a band-pass-filtered part of a supercontinuum generated by the local FC, which is centered around 1145 nm with a 6-nm (45-cm$^{-1}$) FWHM bandwidth and a 2-μW average power. CCS is achieved through SFG of the target FC and the local FC in a 1-mm-long periodically poled lithium niobate (PPLN) crystal followed by its interference with the readout FC, which is measured by a 100-MHz InGaAs balanced detector (Thorlabs PDB415C). The

PPLN crystal (Covesion MOPO1-0.5-1) has a 29.52-μm poling period that can provide a ~200-cm$^{-1}$ (~350-nm) quasi-phase-matching 3-dB bandwidth for the SFG. Supplementary Section 1.1 presents the detailed setup and optical spectra of those FCs.

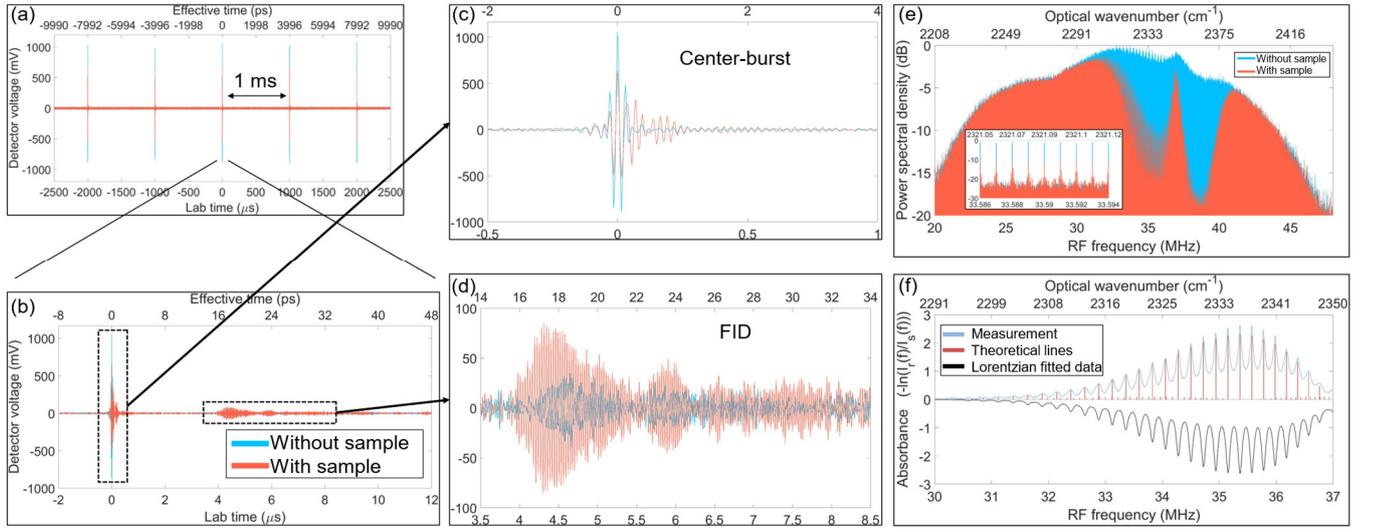

**Fig. 3 Experimental results of CCS of $CO_2$. a** Five consecutive interferograms with a 1-ms temporal spacing, corresponding to $\delta$=1 KHz. The "without sample" result (blue) is measured when the optical path is purged with nitrogen ($N_2$), and the "with sample" measurement (red) is taken when the path is not purged and atmospheric $CO_2$ is present. All measurements are carried out at room temperature and atmospheric pressure without extra control **b** The central 14 μs of one example interferogram. Blowups depicting additional details of the center-burst and FID are shown in panel **c** and **d**, respectively. The lower temporal axes denote the lab time while the upper ones denote the effective time[26], which are related by the equation $t_{Lab}/t_{Effective} = f_{rL}/\delta$. **e** Spectra of band B of the RF FC, obtained by Fourier transforms of 498 consecutive unapodized interferograms, for measurements both with and without $CO_2$, are shown in red ($I_s(f)$) and blue ($I_r(f)$), respectively. The inset is a zoomed-in view to show resolved comb lines, which are separated by $\delta = 1000\ Hz$ in the RF domain corresponding to $f_{r,L} = 250{,}250{,}820\ Hz$ in the optical domain. **f** Measured molecular absorbance spectrum (light blue curve), $A(f)$, defined by $A(f) = -\ln[I_r(f)/I_s(f)]$. The result is obtained from 498 interferograms (for both "with" and "without sample") each apodized with 100-μs window. The black curve (inverted) denotes the theoretical model, which is derived by fitting the absorption lines from the HITRAN database (red lines) with a Lorentzian lineshape to the experimental result. The upper axes in both (e) and (f) denote the optical frequency in wavenumber.

**Experimental results.** Figure 3a presents five consecutive interferograms with a period of 1 ms, out of which the central 14 μs of one interferogram is depicted in Fig. 2b. The prominent effects due to $CO_2$ can be observed in both the center-burst (Fig. 3c) and the tail (Fig. 3d), which is the result of the coherent addition of molecular FID[11,27]. Note that, thanks to the temporal gating, the background power at the tail (FID) is much weaker than that at the center-burst. This background is not visible in the measurement shown in Fig. 3 as it is concealed by the balanced detector, but it is prominent if the detector is not well balanced (Supplementary Fig. 1-3). Based on the measurement, we estimate a single-shot time-domain SNR of 167 ($\frac{\pm 1000mV}{\pm 6mV}$) in a 28-MHz electrical bandwidth, which is more than four times that of a recent EOS work[21]. On the other hand, we estimate an upconversion (SFG) efficiency of at least 2% $mm^{-1}$, which is more than two orders of magnitude higher than that of a recent C.W. upconversion DCS work (Supplementary Section 1.3).

Figure 3e represents the results in the frequency domain, obtained by the Fourier transform of 498 consecutive interferograms (498 ms) without apodization for both "without sample" and "with sample" cases, where ~2.78×10$^4$ comb teeth are present in a 245-cm$^{-1}$ band. The average SNR of the without-sample spectrum is 28.9, which gives a sum of spectral SNR of 8.03×10$^5$ (the sum of the SNR of all comb teeth). Note that we are only able to acquire 0.5-s data with $\delta$=1kHz, which means the signal spectrum only uses 28 MHz of the whole 125-MHz Nyquist band (half $f_{rL}$), limited by the memory depth of our data acquisition equipment. If we use a factor of 125/28 larger $\delta$ and acquire for 1s, we can get ~9 times as many interferograms, which will scale up the sum of spectral SNR by a factor of 3. This leads to an estimated figure of merit of $2.4\times10^6$ Hz$^{1/2}$ for this MIR spectrometer (Supplementary Section 1.3), which is one of the highest among recently reported MIR DCS or EOS works[4,5,7,21,24]. Note that our SNR can still be further increased since the time-domain signal only reaches about half of the detector saturation ($\pm$ 1V out of $\pm$ 1.8 V), which can be increased through a higher power from any of the target, local or readout FC.

Shown in Fig. 3f (light blue curve), the molecular absorbance spectrum is obtained by comparing with-sample and without-sample measurements (here a 100-μs apodization window is applied to the interferograms before FFT). Only the *P* branch (rotational structure below the band origin) of the measured spectrum of $CO_2$ is shown here (see Supplementary Section 1.4). The theoretical absorbance spectrum (black curve, inverted about the x-axis for clarity) is calculated using spectral lines (red lines) from the HITRAN database[28] fitted with a Lorentzian line shape of 0.8-cm$^{-1}$ FWHM linewidth.

Note that these results are obtained by locking the $f_{rep}$ of the target and local combs only individually to a RF standard, which gives fixed $f_{rep}$ values but an uncontrolled broad relative linewidth between two combs[29]. By some post-processing (without external optical referencing), we can correct the without-sample measurements but are unable to fully correct the with-sample measurements (Supplementary Section 1.4). We believe this is the main reason for why the fitted absorption linewidth (0.8 cm$^{-1}$) is larger than the theoretical pressure broadening (~0.2 cm$^{-1}$) at room temperature and atmospheric pressure. Similar to other DCS techniques, this problem can be solved by utilizing an intermediate C.W. reference to provide fast phase noise information for either tight-locking using fast actuators[4–7,11,30] or error correction by post processing[5,29,31–33].

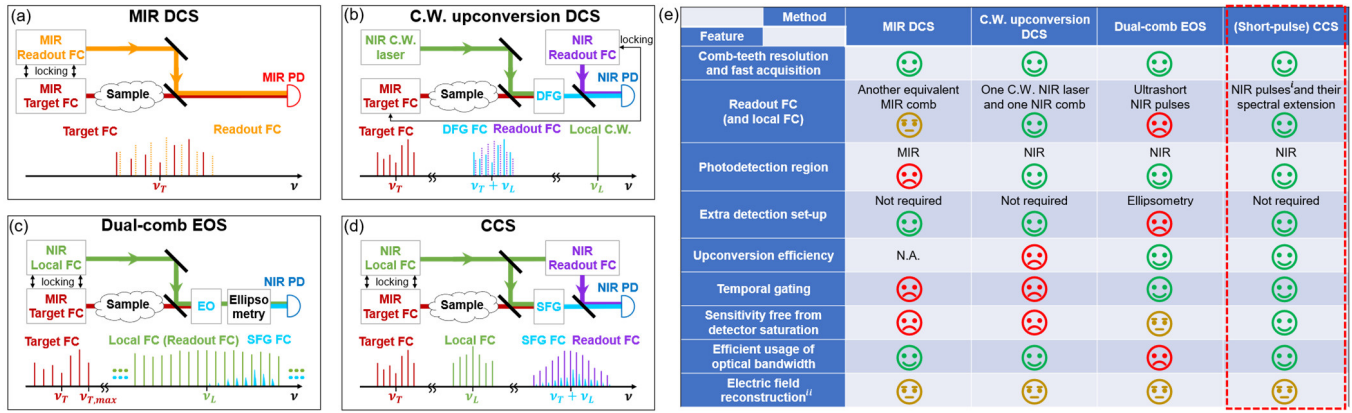

**Fig. 4 Comparison of principles and features of different dual-comb-based techniques for MIR spectroscopy. a-d** Simplified schematics of different techniques. **a** General dual-comb spectroscopy with an asymmetric (dispersive) configuration[1]. The second MIR FC, which does not pass through the sample, is often referred to as the "local FC" or "slave FC" in other works. However, in the context of this work, it is named as "MIR readout FC" since it samples the MIR target FC linearly, by which a linear cross-correlation signal is generated to give the spectral information of the target FC. **b** C.W. upconversion DCS. The MIR target FC is generated by the DFG between the NIR C.W. laser and the "master NIR comb"[24], which is not shown in this simplified schematic. This method can be considered as a special case of CCS, in which the "local FC" contains only one "comb tooth". Note that using an SFG or DFG process for the nonlinear upconversion of the MIR target FC does not make a fundamental difference. **c** Dual-comb EOS. It can also be considered as a special case of CCS, in which the local FC is so broadband that it also serves as the readout FC. The lower-frequency part of the local FC can be regarded as an effective "local FC", while the higher-frequency part can be regarded as an effective "readout FC", in the context of CCS. **d** General cross-comb spectroscopy. **e** Table of a comparison of features of different techniques. (*i*) In principle, CCS does not require short pulses as the NIR local FC. However, the (short) local pulses enhance the upconversion efficiency and enable temporal gating. (*ii*) If electric field of the readout FC (and local FC if applicable) is (are) known, all four techniques can fully reconstruct the electric field of the target pulse. However, this extra information is not necessary for the purpose of general absorption spectroscopy.

## Discussion

CCS has many advantages compared to other dual-comb-based techniques despite their similarities. Figure 4a-d illustrates dual-comb-based spectroscopic techniques in the MIR, including DCS, C.W. upconversion DCS, dual-comb EOS, and the CCS. In all four techniques, two combs of slightly detuned repetition rates are employed to replace the mechanical scanning stage used in traditional FTIR techniques. In fact, CCS can be considered the general form of frequency-converted DCS. C.W. upconversion DCS and dual-comb EOS are essentially two special cases of the CCS; the former uses a very narrow-band local "FC" with only one "comb tooth", and the latter uses a very broadband local FC (very short local pulses) which also functions as the readout FC (Supplementary Section 3). The features of these four techniques are summarized in Fig. 4e. Compared to MIR DCS, CCS is free from requiring a second MIR FC as well as the poor performance of MIR detectors, and it can have enhanced performance from temporal gating. Compared to C.W. upconversion DCS, CCS features much higher upconversion efficiency and temporal gating, thanks to the short local pulses. Compared to EOS, CCS does not require ultrashort sampling pulses and ellipsometry to detect polarization rotation, which can be experimentally complex and challenging, and it can fully avoid the limitation from detection saturation, so it is more resource-efficient and flexible.

The SNR and sensitivity of the CCS are limited by the upconversion efficiency, instead of the detector and background noise. There are three factors in the upconversion process: the target pulse (generally MIR), the local pulse (generally NIR), and the nonlinear platform (generally a bulk crystal). The improvement of any of these elements can be exploited to improve the performance and capability of CCS. First, in the past decade, there has been impressive progress in the high-power (>100 mW) MIR FCs [14–20]. However, MIR DCS cannot take full advantage of this progress, as MIR detectors typically saturate at ~1 mW. One can always apply more detectors and bandpass filters to do parallel and sequential detection[12] to alleviate this problem, which adds to the system complexity and cost, and still does not address the limitation from the strong background noise at the FID due to the nature of linear interference. Conversely, CCS can reap the full benefits of more powerful target pulses as saturation is no longer a problem. Although the target pulse used in our experiment is already among the FCs with the highest power in its wavelength region, it can still be improved[14] and enhance the CCS performance.

Second, for the local FC, we use 1.56-μm local pulses with only a 200-mW average power and 100-fs pulse width, while near-IR combs with orders of magnitude better metrics are already available and CCS can directly benefit from them. Third, a 1-mm commercial bulk PPLN crystal is used in our experiment as the nonlinear medium, which limits the upconversion efficiency and bandwidth. Recently, developments in lithium niobate nanophotonics have enabled dispersion-engineered waveguides with unprecedented phase-matching bandwidths and nonlinear efficiencies orders of magnitude higher than bulk PPLN crystals[34,35], which can tremendously improve CCS performance.

Although the upconversion efficiency demonstrated in our experiment is still low, the NIR detector is already half saturated at the center-burst. The detector can be fully saturated by simply doubling the power of local FC or readout FC. By combining part or all these above-mentioned state-of-the-art techniques, we expect the FID signal from trace molecules can get close to the saturation limit of the NIR detector so that CCS can provide record high-SNR mode-resolved spectroscopy for record low concentration samples. This is expected to break the limit of current dual-comb-based techniques and is highly desirable for applications where broadband trace molecular detection plays a significant role, such as breath analysis.

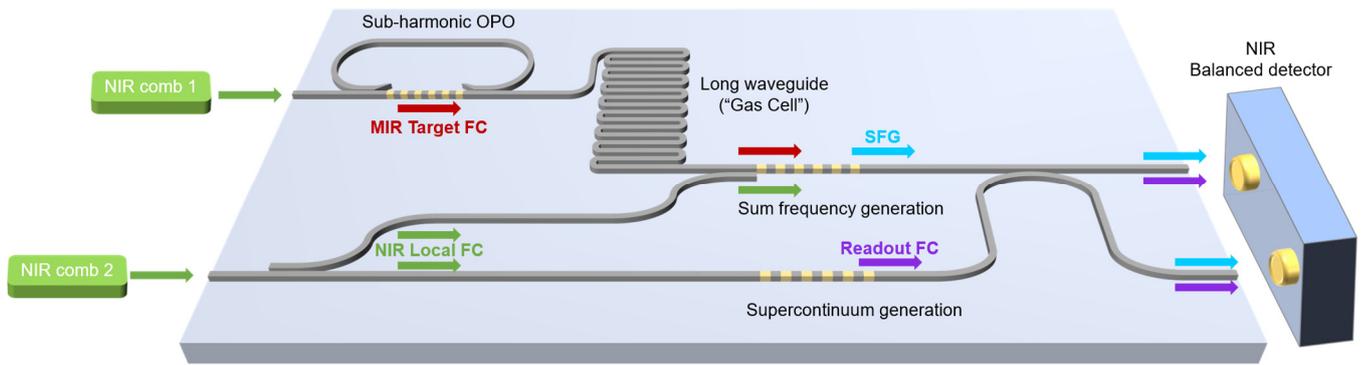

**Fig. 5 Envisaged on-chip implementation of CCS.**

Currently, our setup relies on free-space and fiber-based components. However, rapid progress in lithium niobate nanophotonics[36,37] has made it feasible to monolithically integrate most of the components of the entire CCS system into a single photonic chip. In Figure 5, we show an envisaged on-chip implementation, in which two NIR combs are coupled to the chip, and one of them is used to pump a sub-harmonic OPO to generate the MIR target FC[38,39]. The other NIR comb plays the role of the local FC in sum-frequency generation[38,40,41], while part of it is used to generate the readout FC via supercontinuum generation[41–44]. A long waveguide is used to increase the interaction area between the target FC and the surrounding environment. Moreover, recent progress in on-chip NIR combs and detectors based on thin-film lithium niobate[36,37] suggests the potential to also integrate both NIR sources and detectors, which can lead to a fully integrated CCS system.

In summary, we introduce the new concept of cross-comb spectroscopy, which can not only convert spectral information to a more easily accessible wavelength region but also overcomes the limits of the general dual-comb spectroscopy. CCS can combine and improve upon many of the merits of other demonstrated techniques while circumventing some of their practical challenges. We experimentally demonstrate a CCS measurement around 4 µm with a broad bandwidth, high SNR, and large figure of merit, which are among the best of reported metrics at this wavelength range. This work opens a simple, flexible, and efficient avenue to high-precision, high-sensitivity, high-SNR, high-speed, and broadband spectroscopy in spectral regions with less developed sources and detectors.

## Acknowledgements
The authors gratefully acknowledge support from AFOSR award FA9550-20-1-0040, NSF Grant No. 1846273, and NASA/JPL. The authors thank Luis Ledezma at Caltech for helpful discussion. The authors thank Zhiquan Yuan, Lue wu, and Prof. Kerry J. Vahala at Caltech for loaning equipment. C.R.M. is grateful for support from the Arnold O. Beckman Postdoctoral Fellowship.

## Author Contributions
The manuscript was written through contributions of all authors.

## Competing Interests
The authors declare no competing interests.

## Data availability
The data that support the plots within these paper and other findings of this study are available from the corresponding author upon reasonable request.

## Code availability

The codes used for Fourier transform are available from the corresponding author upon reasonable request.

# Supplementary Materials for

# Mid-Infrared Cross-Comb Spectroscopy

**Mingchen Liu[1], Robert M. Gray[1], Luis Costa[1], Charles R. Markus[2], Arkadev Roy[1], Alireza Marandi[1, *]**

[1]Department of Electrical Engineering, California Institute of Technology, Pasadena, California, 91125, USA
[2]Division of Chemistry and Chemical Engineering, California Institute of Technology, Pasadena, California, 91125, USA
[*]marandi@caltech.edu

# Table of contents

# 1. Setup and supplementary experimental results

## 1.1 Setup and optical spectra

The setup diagram is depicted in Supplementary Fig. 1-1, and the optical spectra are depicted in Supplementary Fig. 1-2.

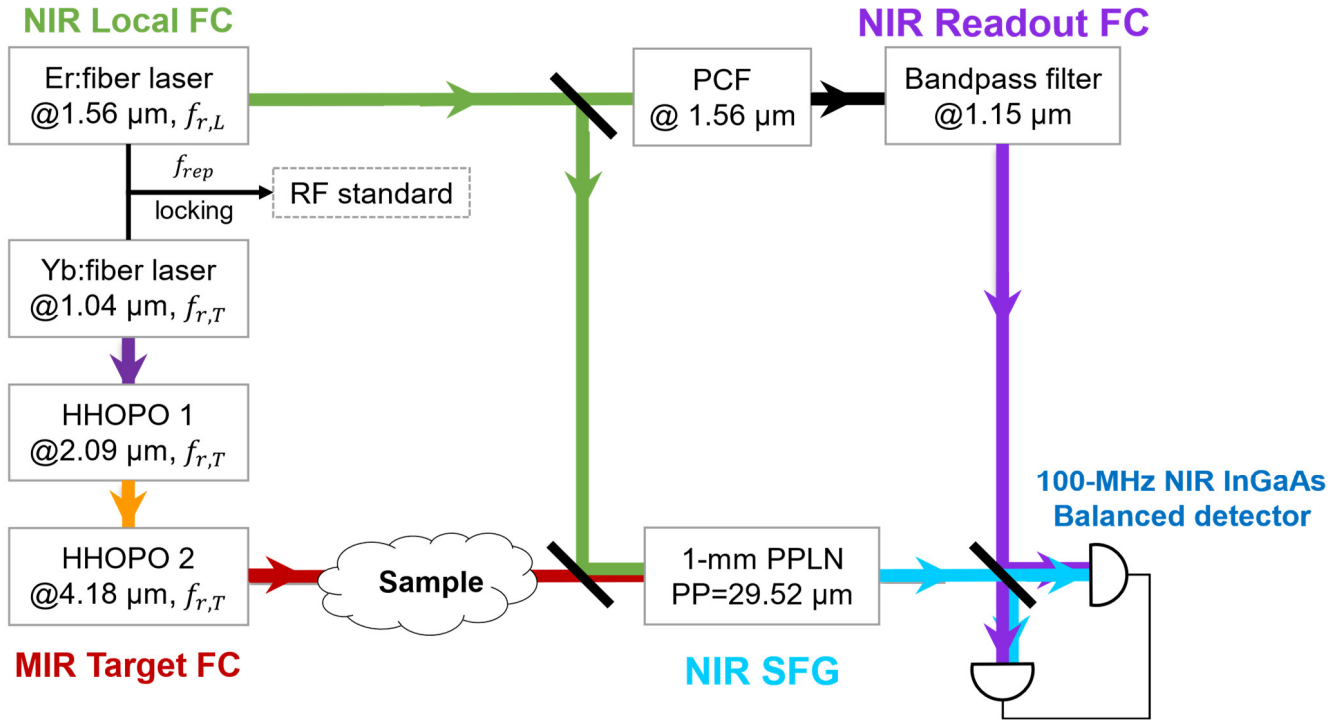

**Supplementary Fig. 1-1| Experimental setup.** PCF: photonic crystal fiber. PP: poling period. Note that the mixing between the SFG FC and readout FC is realized in a 50:50 fiber coupler, before which the two beams are coupled from free space into fiber. The configuration of fiber coupler is further illustrated in Supplementary Fig. 3a. The bandpass filter is centered at 1140nm with a full width half maximum (FWHM) of 15 nm.

The target FC is provided by a chain of two cascaded half-harmonic OPOs[1]. Pumped by a commercial mode-locked Yb: fiber laser centered at 1.045 μm, the first half-harmonic OPO generates 2.09-μm pulses, which are then used to pump the second half-harmonic OPO at 4.18 μm. Half-harmonic OPOs feature intrinsic phase and frequency locking of their output to the pump[2]; thus the phase and frequency of the 4.18-μm OPO are intrinsically locked to that of the 1.045 μm pump. Hence, by locking the $f_{rep}$ ($f_{ceo}$) of the 1.045-μm laser to that of the 1.55-μm Er: fiber laser (local FC), the target FC (4.18-μm OPO) is locked to the local FC. In this experiment, the $f_{rep}$ and $f_{ceo}$ of the local FC and 1.045-μm Yb fiber laser (target FC) and all measurement apparatus are locked to a 10-MHz RF rubidium (Rb) clock, ensuring a common frequency standard.

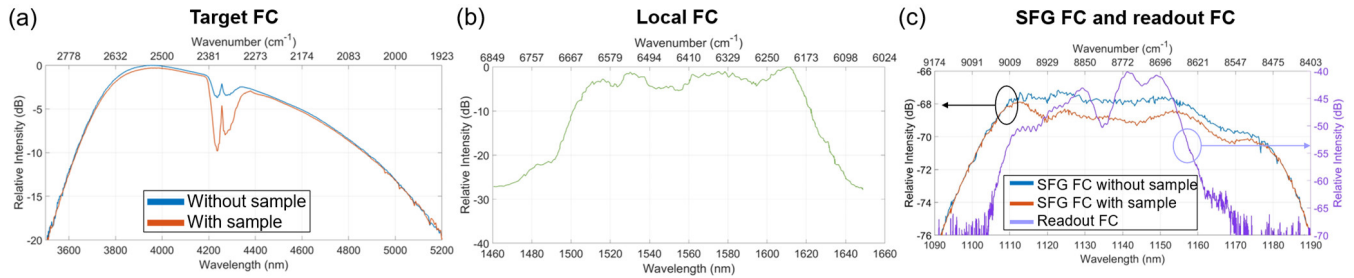

**Supplementary Fig. 1-2| Optical spectra of frequency combs used in the experiment. a.** Target FC for both "without sample" (purged) and "with sample" (unpurged) cases, measured by a commercial Fourier-transform infrared spectrometer (FTIR) with a resolution of 4 cm$^{-1}$. The residual $CO_2$ which cannot be fully cleared by purging is the reason why the absorption dip can still be observed in the "without sample" curve. **b.** Local FC spectrum provided by the manufacturer (Menlo Systems). **c.** Spectra of SFG FCs (with and without sample) and readout FC measured by a grating-based OSA with a resolution of 0.5 nm.

*1.2 Temporal gating*

For the same reason as in EOS[3], our cross-comb method can also benefit from the temporal gating (also referred to as "nonlinear gating"), although our local pulse is not as short as that of EOS. However, this effect cannot be seen in the measurement shown in Fig. 3 of the main paper because the balanced detection conceals the strong background. Corresponding to band A and band D of the RF FC (see Section 2.4), the background is a common-mode signal only from the port of the SFG FC and thus is cancelled by the balanced detection. Note that the background at the center-burst is basically an intensity cross-correlation of the target pulses and local pulses, so it is delay ($\tau$, lab-time)-dependent unlike DCS (See Fig. 1 of the main paper). As shown in Supplementary Fig. 1-3, if we tweak the coupling of the splitter to the balanced detector (panel (a)) such that it is not well-balanced, the strong background will show up prominently at the center-burst (panel (b)). However, because of the temporal gating, the beating at the tail, which contains useful information, is free from any undesirable background power from the center-burst. The complete description and importance of temporal gating is more complicated and can be found in section 3 and 4.2.

Note that the balanced detection can only "conceal" the background in its RF output, but it cannot solve the problem caused by the strong background. Although a well-balanced detector can cancel the common-mode signal and noise in its RF output by comparing the outputs of two photodiodes, there may still be strong common-mode optical power incident on each photodiode which is not visible in the balanced output. The strong incident optical power can bring in noise which is not common-mode and thus cannot be cancelled, and it will ultimately saturate the photodiodes. This problem exists in detection of weak FID signal for both DCS and EOS.

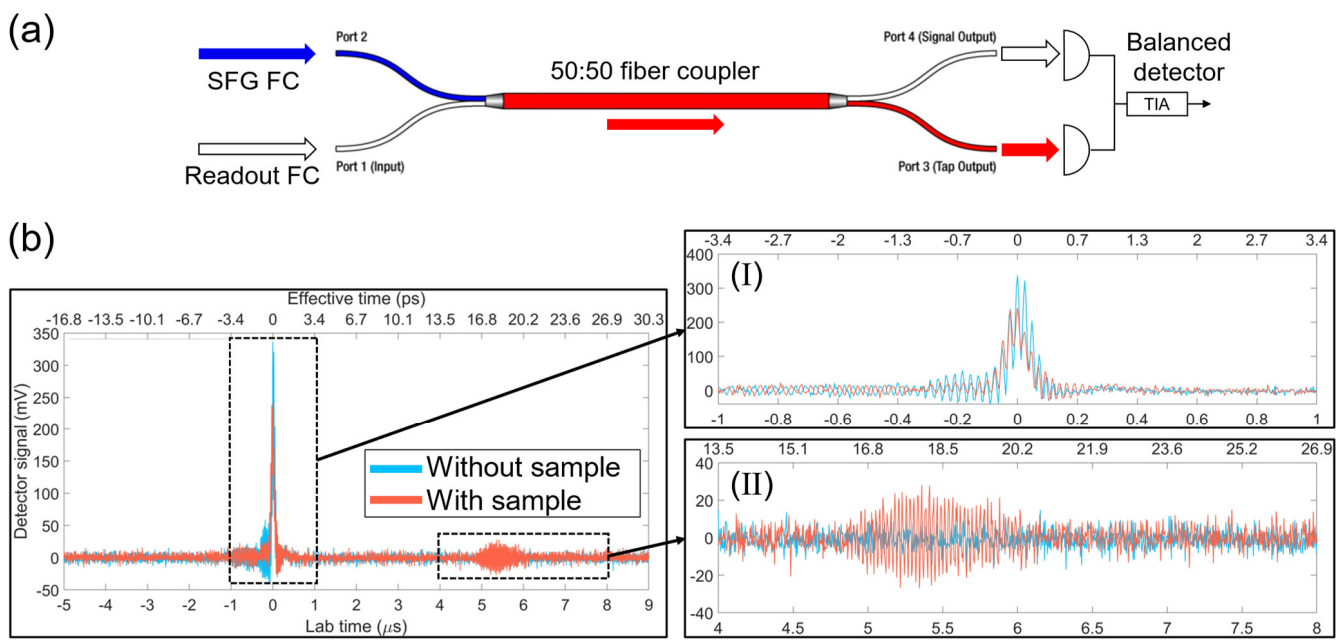

**Supplementary Fig. 1-3| Interferograms of CCS of $CO_2$, measured by an unbalanced detector. a.** Configuration of the fiber coupler and balanced detection. TIA: transimpedance amplifier. **b.** Interferograms measured when the detector is not well balanced. The main figure presents the central 14-µs part of one example interferogram in order to highlight the details of the center-burst (inset I) and the tail (inset II). Note that the measurement is done when the detector is tuned to be just slightly unbalanced. The background at the center-burst is actually very strong and can heavily saturate the TIA if the detector is further unbalanced. Note that the result shown here is from an older measurement where local pulses with lower power are used; thus, its FID signal is lower than that of the Fig. 3 of the main paper.

*1.3 Estimation of experimental signal-to-noise ratio (SNR) and figure of merit (FOM)*

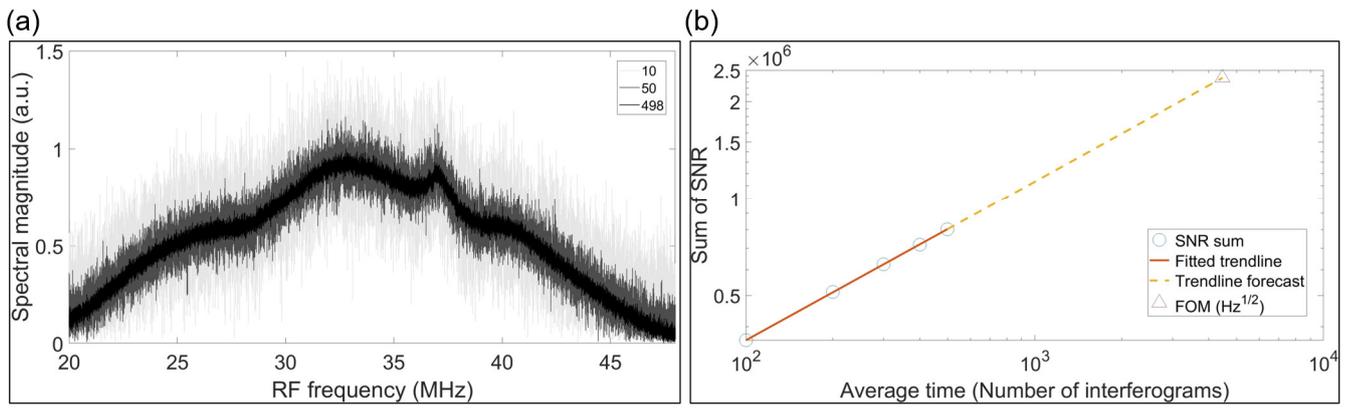

**Supplementary Fig. 1-4| estimation of SNR and FOM. a,** averaged without-sample spectra by different numbers of interferograms (10,50,498, denoted by curves with different grayscales). **b.** Sum of SNR as a function of average time (N), where N denotes the number of averaged interferograms. The red line is a linear fitting of the data points (blue circle), and the yellow dashed curve denotes the forecast the trendline, which scales the experimental sum of SNR to an estimation of the FOM (purple triangle). Note that the coordinate is in log-log scale.

Supplementary Fig. 1-4 presents the estimation of SNR and FOM of our spectrometer. Panel (a) shows averaged without-sample signal spectra by different numbers of interferograms. Note that a small part of the FID signal in the interferograms is discarded before the Fourier transform is applied to exclude the influence of residual sample absorption. It is readily seen that the SNR of the signal spectrum increases with the averaging time. Panel (b) depicts the spectral SNR sum (the sum of the SNRs of all spectral components) as a function of N (number of averaged interferograms) on a log-log scale, with experimental data denoted by the blue circles. A linear fitting (red line) is conducted for those experimental points, whose slope is 0.4936, indicating the SNR increases as a function of $\sqrt{N}$ as expected.

Because of the reasons mentioned in the main paper, currently we are only able to acquire ~0.5-s data with a $\delta$ of 1 kHz, which gives a SNR sum of $8.03\times10^5$ (the highest blue point). To estimate a figure of merit (FOM), we need to scale the number in two ways. Firstly, we assume data acquisition over a full second, which can give twice as many interferograms. Secondly, the signal currently only takes up 28 MHz out of the whole available 125-MHz spectrum (half repetition rate of the local/target FC). If we could set the $\delta$ to be ~4.5 kHz (we did not do that in the experiment due to some limitations of our detector and locking electronics), we would be able to get ~4.5 times more interferograms. In total, we can realistically obtain 9 times as many interferograms in our measurement, which leads to an estimation of FOM of $2.4\times10^6$ (yellow dashed line and purple triangle).

In addition, here we explain how we estimate the upconversion efficiency of our experiment and how it compares to that of prior work using C.W. upconversion[4]. In that work, a 700-µW MIR FC and a 2.4-W C.W. laser are used to get a 1.3-µW upconverted NIR FC with a 20-mm PPLN crystal. Note that all the power in this subsection refers to average power. Therefore, their upconversion efficiency can be calculated as:

$$\frac{P_{upconverted}}{P_{MIR} \times L_{crystal}} = \frac{1.3\ \mu W}{700\ \mu W \times 20\ mm} = 9.3 * 10^{-3}\ \%\ (mm^{-1})$$

In our experiments, we use 500-mW target pulses and 200-mW local pules to generate a SFG signal of ~100 nW with a 1-mm PPLN crystal. However, in our case, the 100-fs local pulse scans through the 50-fs target pulse due to their different repetition periods (see Fig. 1e of the main paper), so the real time interval in which the two pulses overlap (when SFG power is generated) only accounts a very small portion of the full period. Specifically, when we set the $\delta = 1\ kHz$ for the $f_{rep,L} \cong f_{rep,T} = 250\ MHz$, the local pulse will scan through the target pulse in a total of $2.5 \times 10^5$ steps with a step size of approximately 16 fs (meaning the relative position of local and target changes by about 16 fs for each local pulse which samples the target) over the course of an interferogram. Since the target pulse is only around 50 fs, there are only 3~4 steps in which the local pulse overlaps with the target pulse well out of those 2.5 x $10^5$ steps, and moreover, there is only at most one step that the two pulses overlap perfectly (the maxima of the interferogram). Therefore, we can estimate an effective overlap coefficient (the "duty cycle" of SFG generation) to be about $10^{-5}$, which must be factored into the calculation of the SFG efficiency for a fair comparison. Therefore, the efficiency is:

$$\frac{P_{upconverted}}{P_{MIR} \times L_{crystal}} \times \frac{1}{duty\ cycle} = \frac{100\ nW}{500\ mW * 1mm} \times \frac{1}{10^{-5}} = \frac{1}{5*10^6} \times \frac{1}{10^{-5}} mm^{-1} = 2\%\ (mm^{-1})$$

Our upconversion efficiency is more than two orders of magnitude higher than the C.W. upconversion work, although the average power of our local FC is just one tenth of theirs. Note that we couple the generated SFG from free space to a single-mode fiber and then measure its power with a fiber-coupled OSA. However, there could be a large loss in the free space-fiber coupling which results in an underestimation of the measured SFG power. Therefore, this efficiency could be correspondingly underestimated.

*1.4 Phase correction and broadened absorption linewidth*

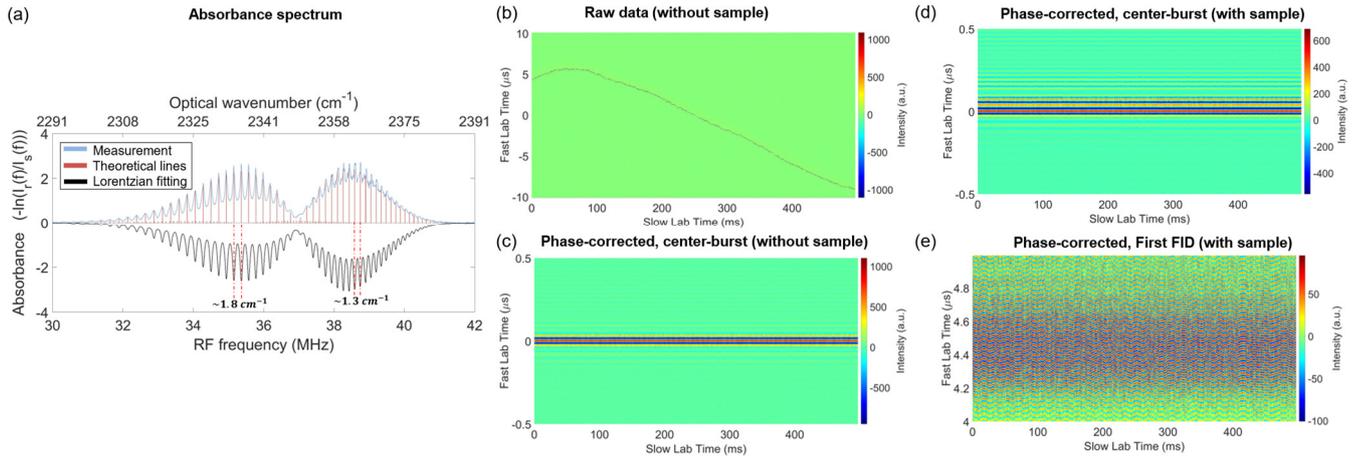

**Supplementary Fig. 1-5|. Full measured absorbance spectrum of atmospheric $CO_2$ and phase correction process. a,** Full measured absorbance spectrum of atmospheric $CO_2$ in our preliminary cross-comb measurement, including both P and R branches. The SNR of the R branch is lower than that of the P branch. Also, the absorption lines of the R branch are broader than that of the P branch. The spacing between absorption lines in R branch is smaller than that of the P branch. **b,** 3D interferogram of the raw data of the without-sample measurement. **c,** 3D interferogram of the corrected data of the without-sample measurement, around center-burst. **d,** 3D interferogram of the corrected data of the with-sample measurement, around center-burst. **e,** 3D interferogram of the corrected data of the with-sample measurement, around first peak in FID at a fast lab time interval of [4,5] μs.

Supplementary Fig. 1-5a shows the full measured absorption spectrum, whose right side (higher optical frequency, R branch) has a worse SNR and more broadened absorption linewidth compared to its left-side counterpart (lower optical frequency, P branch). This is because the phase noise (uncontrolled broad relative linewidth) between the target FC and local (readout) FC has a larger effect on the R branch, as explained below.

Panels (b)-(d) show the phase correction for the interferograms, where "3D interferograms" are presented. In those 3D interferograms, each column is a single interferogram (detector voltage is denoted by the colormap), and consecutive single interferograms (columns) are plotted from left to right. Therefore, the vertical axis denotes the "fast time" within each single interferogram, and the horizontal axis denotes "slow time" that shows the time spacing between each interferogram.

Panel (b) shows the without-sample 3D interferogram with the fast time zoomed in to [-10,10] μs to show the center-burst. Ideally, the center-bursts of each single interferogram should perfectly align at $t_{fast} = 0$, but they shift due to the phase noise and timing jitter between the two fiber lasers since no tight locking is applied to them[5]. Two steps are taken to correct these shifts. Firstly, the maxima of the envelopes of each interferogram are shifted and aligned to correct the timing jitter (the envelope of the interferograms is obtained by Hilbert transform). Secondly, a zero-order phase term is applied to each interferogram to make its phase at the maxima of the envelope zero to correct the zero-order phase shift between interferograms.

The obtained results are shown in panel (c)-(e). At the center-burst of both without-sample and with-sample case (panel (c) and (d)), the corrected interferograms overlap very well and thus can be coherently averaged. This is largely because our correction uses the information from the sharp peak structure at the center-burst of the interferograms and thus provides reasonably good correction over the whole center-burst, since it is generated by the interference of two short, femtosecond pulses. This is sufficient for the without-sample measurement since its information only exists around the center-burst. However, for the with-sample case, although its center-bursts are aligned well and can be averaged, prominent phase error still exists at larger fast times, for example, at the first strong FID peak (panel (e)). This is because the coherence time between the two combs is smaller than 4 µs, so the zero-order phase correction at the center-burst is not able to fully correct the error at the more distant FID. The larger the fast time (delay $\tau$), the larger the phase error between different interferograms, and the less they can be averaged. In other words, the relative comb linewidth between our target and local (readout) combs remains broad because their repetition rates are independently locked only to a RF standard.[5]

This explains why our measured absorption linewidth is larger than the theoretical value; although the center-burst of the with-sample measurement can average well, the FID signal at large delay $\tau$ cannot. This is like a window function is applied to the time-domain of the averaged with-sample interferogram, which is equivalent to a sinc function convolved to its spectrum which broadens all its spectral features. The R branch of the absorption spectrum suffers more from this undesirable effect because its corresponding time-domain information exists at a larger time delay (fast time), which means a larger phase error and less averaging due to its smaller frequency domain spacing between different absorption lines. As mentioned in the main text, this issue can be solved by setting one intermediate C.W. reference to provide information for either tight-locking by fast actuators, or error correction by data processing, as has been well demonstrated in dual-comb spectroscopy.

However, this effect does not influence the without-sample measurements since their information only exists close around the center-burst in the time-domain where phase error and timing jitter can be corrected based on the information from the sharp peak structure of the center-burst itself. Therefore, in terms of SNR, the obtained without-sample measurements after our phase correction are comparable to the results that can be obtained if the relative comb linewidth (phase noise and timing jitter) between target and local combs are ideally controlled. Hence, our estimate of the FOM of our spectrometer using the SNR of the without-sample measurements is fair.

## 2. One-to-one tooth mapping from target FC to RF FC

### 2.1 Target FC and Local FC

The electric field of the local FC can be described by

$$e_L(t) = \sum_m A_m^L \exp(i\phi_m^L)\exp(-i2\pi\nu_m t) = \sum_m L_m \exp(-i2\pi\nu_m t)$$

where $L_m$ denotes the complex amplitude that encodes both the intensity and phase of the $m^{th}$ local comb tooth with optical frequency $\nu_m$, and the spatial dependence is omitted here. The superscript "L" of $A_m^L$ and $\phi_m^L$ denotes local FC, and the subscript "m" corresponds to the $mth$ comb tooth. In addition, for the optical frequency $\nu_m$, we have

$$\nu_m = m f_{r,L} + f_{ceo,L}$$

where $f_{r,L}$ and $f_{ceo,L}$ are the repetition rate and carrier-envelope offset (CEO) frequency of the local FC, respectively.

Sometimes it is not convenient to directly use "$m$" to index comb teeth, since the first tooth usually occurs at very large m. To be specific, for the first tooth of a practical frequency comb, $m_{first} \sim 10^6$. For convenience, here we define the effective tooth index, $m'$, which starts at 1. If we use $m_{first}$ to denote the tooth index of the first local tooth, we have:

$$\nu_{m'=1} = \nu_{m_{first}} = (m_{first} - 1)f_{r,L} + 1 \times f_{r,L} + f_{ceo,L} = 1 \times f_{r,L} + f_{ceo,L} + \nu_{start,L}$$

Where $\nu_{startL} = (m_{first} - 1)f_{rL}$.

Then, for any $m'^{th}$ teeth ($m'$ starts from 1),

$$\nu_{m'} = m' f_{r,L} + f_{ceo,L} + \nu_{start,L}$$

Like the local FC, we use $X_n$ to denote the complex amplitude of the $n^{th}$ target comb tooth with the optical frequency $\nu_n$. We have

$$e_T(t) = \sum_n X_n \exp(-i2\pi\nu_n t), \quad \nu_n = n(f_{r,L} + \delta) + f_{ceo,T}$$

Where $\delta$ denotes the repetition rate detuning between the local and target FC, i.e., $f_{r,T} = f_{r,L} + \delta$.

Similarly, we can also define the effective tooth index for target comb teeth,

$$\nu_{n'} = n' f_{r,T} + f_{ceo,T} + \nu_{start,T}, \quad \nu_{start,T} = (n_{first} - 1)f_{r,T}$$

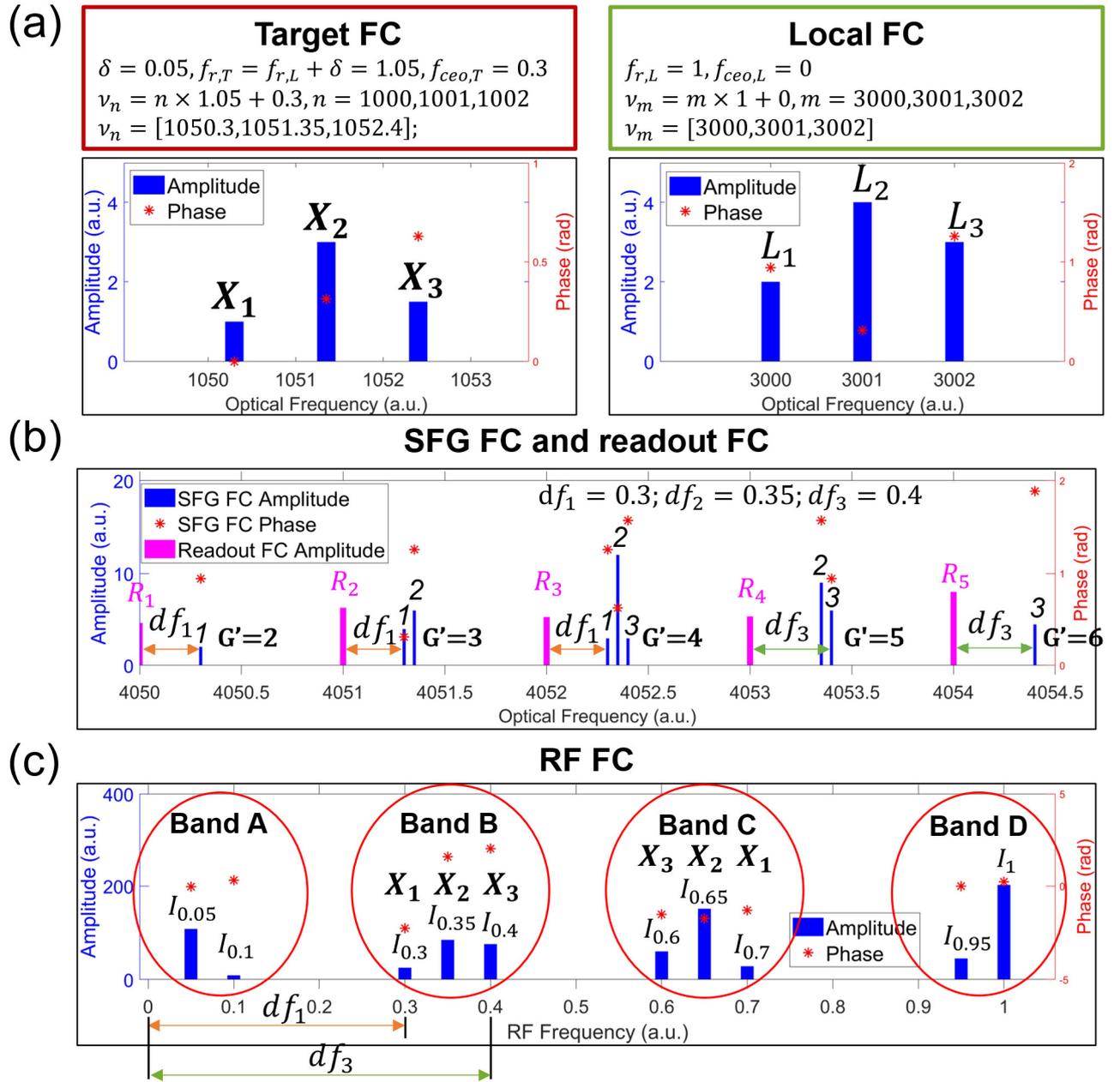

**Supplementary Fig. 2-1| Quantitative illustration of one-to-one tooth mapping of cross-comb spectroscopy. a.** Target FC and local FC. Effective tooth indices are used to label each comb tooth in the plot. **b.** SFG FC and readout FC. Each SFG tooth is labeled by the effective tooth index ("1", "2", or "3") of the corresponding target tooth. The phase for the readout FC is assumed to be constant for each tooth and is thus not shown in the plot. $df_{n'}$ denotes the primary readout frequency for the $n'^{th}$ (effective tooth index) target tooth. Each SFG group is labeled by its effective group index G'. **c.** RF FC. Every RF comb tooth in band B and band C is labeled with its corresponding target tooth.

Using the notation introduced above, the frequency-domain picture of cross-comb spectroscopy is depicted in Supplementary Fig. 2-1. To make a concise and clear illustration, only three teeth are included for both FC, and simple random numbers are assigned to their optical frequencies, which are of arbitrary unit (Supplementary Fig. 2-1a). Note that generality is not lost by assigning $f_{ceo,L} = 0$, since in practice it is just the relative $f_{ceo}$ between the two FCs that matters. Although only a small number of comb teeth and simple random numbers are used for the following illustrations and equations, the conclusions still hold when scaled to practical numbers.

## 2.2 SFG FC

Because of the slightly detuned repetition rates between the local and target FCs, each pair of teeth from them will generate an SFG tooth at a unique frequency, the set of which are referred to as the SFG FC. The electrical field of a certain SFG tooth can be described by (phase-matching effect is not included here)

$$E_{n,m}^{sfg} = L_m X_n \exp(-i2\pi \nu_{n,m}^{sfg} t), \nu_{n,m}^{sfg} = \nu_n + \nu_m$$

As shown in Supplementary Fig. 2-1b, the resultant SFG comb teeth cluster into different frequency groups[6], which can be indexed by the group index $G = n + m$ (or effective group index $G' = n' + m'$), and the groups are evenly spaced by $f_{r,L} = 1$. The group $G'$ is generated by the SFG between the ... $(n'-1)^{th}, n'^{th}, (n'+1)^{th}$ ... target teeth and the ... $(m'+1)^{th}, m'^{th}, (m'-1)^{th}$ ... local teeth. Note that the center group, with $G' = 4$, contains information about all the target teeth, in spite of the fact that different target teeth are modulated by different local teeth (see also Fig. 1 of the main paper). Such a group that contains the information for all target teeth is called a "complete (SFG) group" in the following context. It is readily seen that the number of complete groups formed is determined by the number of local teeth relative to target teeth.

More patterns can be observed within SFG groups. Firstly, SFG teeth in a single group are separated by $\delta$. Secondly, mixing with different local teeth, a given target tooth will generate multiple SFG teeth, which are all at the same relative frequency position in their respective SFG groups. To illustrate the second pattern, each SFG tooth in Supplementary Fig. 2-1b is labeled by its corresponding target tooth ("1", "2", or "3"). The pattern is made clearer still if readout teeth are introduced as frequency references (see next subsection). These patterns make it possible to do one-to-one mapping between the MIR and RF domains.

## 2.3 Readout FC

To read out the spectral information of the target FC contained in the SFG FC, another comb, referred to as the readout FC, is employed to beat with the SFG FC on a square-law photodetector. The readout FC is effectively a spectral extension of the local FC and therefore inherits its $f_{rep}$ and $f_{ceo}$. As shown in the Supplementary Fig. 2-1b, readout comb teeth can be regarded as "boundary markers" for SFG groups, since they share the same constant distance $f_{r,L}$ between each unit. For a certain SFG group, we name its closest (second closest) readout tooth as its "primary (secondary) readout tooth". For a certain SFG tooth within a SFG group, we name the frequency difference between the tooth and its primary (secondary) readout tooth as its "primary (secondary) readout frequency", and the sum of its primary and secondary readout frequencies is $f_{r,L}$. As shown in the illustration, the SFG teeth generated by the same target tooth always have the same primary readout frequency, even though they are distributed in different SFG groups and correspond to different primary readout teeth. Also, SFG teeth generated by different target teeth have different primary readout frequencies, denoted by $df_{n'}$ in the illustration. These two patterns are very important and provide the foundations for the one-to-one mapping.

As with the local and target FCs, we use "$R_q$" to denote the complex amplitude of the $q^{th}$ comb tooth of the readout FC.

$$e_R(t) = \sum_q R_q \exp(-i2\pi \nu_q t), \nu_q = q f_{r,R} + f_{ceo,R}$$

Also, we can define the effective tooth index for readout comb teeth:

$$\nu_{q'} = q' f_{r,R} + f_{ceo,R} + \nu_{start,R}, \nu_{start,R} = (R_{first} - 1) f_{r,R}$$

Note that $f_{ceo,R} = f_{ceo,L}$ and $f_{r,R} = f_{r,L}$.

## 2.4 RF FC, one-to-one mapping, and absorption spectrum

Based on the SFG and readout comb teeth in the optical domain, one can calculate the resultant RF spectrum detected by a single square-law detector. The bandwidth of the detector is assumed to be "1" ($f_{r,L}$), which means that the highest RF frequency the detector can detect is the repetition rate of the local FC, $f_{r,L}$. This is a common condition for many works in dual-comb spectroscopy. To calculate the RF signal (photocurrent) at a given RF frequency, one must sum the contributions from all the comb tooth pairs that can generate heterodyne beating at this frequency.

$$I_{f_0} = \sum_{f^{rf}=f_0} A_1 A_2^*, f^{rf} = v_1 - v_2 = f_0$$

$A_1$ and $A_2$ denote the complex amplitude of the two involved comb teeth, which can be from the SFG or readout FC. The RF frequency of the beating signal, $f_{rf}$, is equal to the difference between the optical frequencies of the two involved comb teeth. Following these equations, for the case of this illustration, the RF signal at different frequencies can be calculated

$$\text{Band A} \begin{cases} I_{0.05} = (L_1 L_2^* + L_2 L_3^*)(X_1^* X_2 + X_2^* X_3) \\ I_{0.1} = (L_1 L_3^*) X_1^* X_3 \end{cases}$$

$$\text{Band B} \begin{cases} I_{0.3} = (L_1 R_1^* + L_2 R_2^* + L_3 R_3^*) X_1 \\ I_{0.35} = (L_1 R_2^* + L_2 R_3^* + L_3 R_4^*) X_2 \\ I_{0.4} = (L_1 R_3^* + L_2 R_4^* + L_3 R_5^*) X_3 \end{cases}$$

$$\text{Band C} \begin{cases} I_{0.6} = (R_4 S_1^* + R_5 S_2^* + R_6 S_3^*) X_3^* \\ I_{0.65} = (R_3 S_1^* + R_4 S_2^* + R_5 S_3^*) X_2^* \\ I_{0.7} = (R_2 S_1^* + R_3 S_2^* + R_4 S_3^*) X_1^* \end{cases}$$

$$\text{Band D} \begin{cases} I_{0.95} = L_3 L_1^* (X_1 X_2^* + X_2 X_3^*) \\ I_1 = (L_2 L_1^* + L_3 L_2^*)(X_1^* X_1 + X_2^* X_2 + X_3^* X_3) + \left( \sum_{q=1}^{4} R_q^* R_{q+1} \right) \end{cases}$$

Note that, for simplicity, here we use subscript effective tooth indices "1,2,3", "1,2,3" and "1,2,3,4,5,6" ($m', n', q'$) to index different comb teeth of the target FC, local FC and readout FC, respectively. The resulting teeth are also illustrated in Supplementary Fig. 2-1c and are referred to as the "RF FC".

As shown in the illustration, RF FC comb teeth can be classified into four bands[7]. Band A consists of the intra-group beat notes, which are generated by two SFG teeth from the same SFG group. Band D is also composed of beat notes generated by two SFG teeth, but the two teeth are from two different adjacent SFG groups. Note that the frequency component with $f^{rf} = f_{rL} = 1$ is a special component in band D which also includes the contribution from beatings between two readout teeth. Band A and band D result from only the SFG FC (excluding $f_{rf} = 1$) and correspond to the envelope of the SFG pulses (cross-correlation signal between target and local FC) in the time domain, which doesn't contain much useful information for our purpose. In contrast, band B, consisting of beat notes between SFG teeth and their primary readout teeth, is a one-to-one mapping of the original target FC. As demonstrated in the equations, the complex amplitude of a certain band B RF tooth is related to and directly proportional to that of only one target tooth, although it is generally modulated by more than one local tooth and readout tooth. Like band B, band C is also a one-to-one mapping of the original target FC, resulting from beating between SFG teeth and their secondary readout teeth. Band B and C contain the exact same information regarding the target FC, which are mirror images of each other, reflected about $f_{r,L}/2$ in the RF domain.

Based on the one-to-one mapping, the absorption spectrum in the MIR region interrogated by the target FC, including both amplitude and phase, can be obtained by comparing the RF band B (C) measured with the sample in the path and the corresponding result measured without the sample in the path (reference).

*2.5 Universality*

In our experiment, we use a MIR synchronously pumped degenerate OPO (centered at 4.18 μm) and Er-doped fiber laser (centered at 1.55 μm) as the target FC and local FC, respectively. The readout FC is a band-pass filtered portion of a supercontinuum pumped by the local FC, which is generated in a photonic crystal fiber (PCF). It should be noted that the scheme of cross-comb spectroscopy (CCS) doesn't have any limitation on the laser techniques used for the frequency comb generation. However, the current implementation benefits from the intrinsic phase locking of the mid-IR comb to the Yb: fiber laser pump. Also, as a special case of CCS, the local FC or readout FC can be replaced by a "frequency comb" with only one tooth, i.e., a C.W. laser. This is explained in depth in the following section.

Moreover, in this derivation, we demonstrate the frequency-up-conversion one-to-one comb tooth mapping by SFG. In fact, it is also possible to realize one-to-one mapping by difference frequency generation (DFG), the derivation of which is very similar. This may be useful in the application of frequency-comb-based spectroscopy in the ultraviolet spectral range or for even shorter wavelengths.

## 2.6 Bandwidth requirements for Local FC and Readout FC

To realize one-to-one mapping for all teeth of the target FC, local FC and readout FC, one must satisfy some requirements which will be discussed in detail in this subsection. To provide a concise discussion, we continue to use the simple illustration above, keeping the number of target teeth to be three but varying the number of local teeth to be 2, 3, or 4. The results are shown in the Supplementary Fig. 2-2. N, M, and Q denotes the number of teeth of the target, local and readout FCs, respectively.

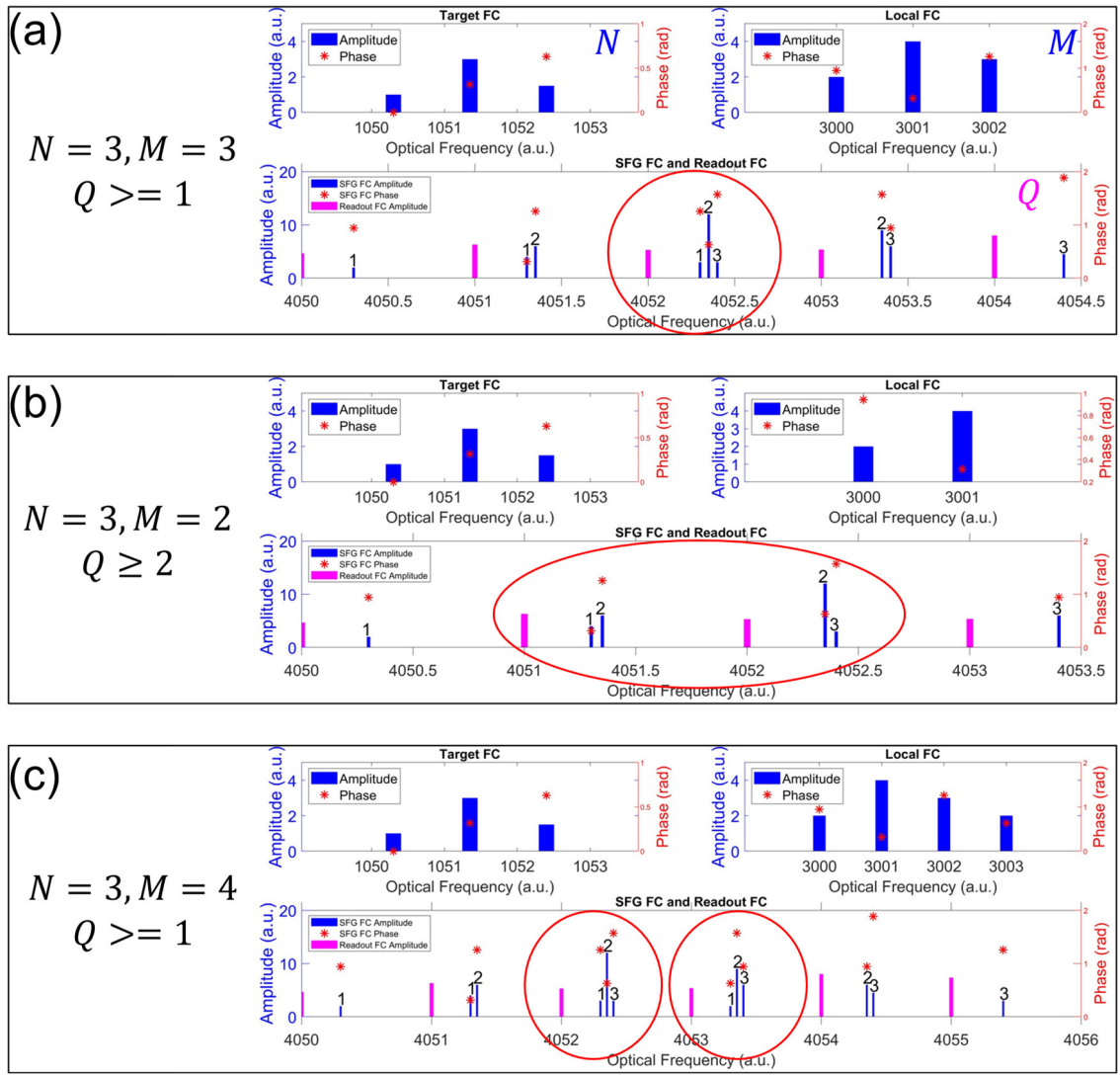

**Supplementary Fig. 2-2| Bandwidth requirements for local FC and readout FC.** M, N and Q denote the number of comb teeth for the target, local, and readout FCs, respectively. **a.** M=N=3, Q must be >= 1. The only complete SFG group, together with its primary readout tooth, is circled in red. **b.** N=2, M=3, Q must be >=2. Two incomplete SFG groups circled in red need to be read out by two readout teeth to map all three target teeth. **c.** N=3, M=4, Q must be >=1. Two complete SFG groups, together with their primary readout teeth, are circled in red.

As shown in the panel (a), when $M = N$, there is only one complete group (circled in red) formed in the SFG FC, which alone contains the information from all target teeth. Thus, to read all target information out, one readout tooth is required at minimum (R>=1), where the equality holds if and only if the readout tooth is the primary (or secondary) readout tooth of that complete group.

If we have one less local tooth (M=2, panel (b)), there is no complete group formed in the SFG FC, and at least two readout teeth are needed to read all three target teeth out (Q>=2). Similarly, to make the equality hold, the readout teeth need to be the primary (or secondary) readout teeth for those two center SFG groups, which are circled in red.

When there is one more local tooth relative to the number of target teeth (M=4, panel (c)), there will be two complete groups (circled in red) formed in the SFG FC. As in the case of L=3, one readout tooth is enough to read out all the target information (Q>=1). However, because of the availability of more complete groups, the requirement of the location of the single readout tooth to make the equality hold is more relaxed compared to the case of M=3. Here, it can be the primary (or secondary) readout tooth of either complete group.

This discussion can be generalized to any large number of teeth, although the various cases are demonstrated only in small numbers here for simplicity. In short, to realize the one-to-one mapping of all target teeth, the minimum required aggregate bandwidth of the local and readout FCs needs to be equal to or greater than that of the target FC, i.e., $M + Q \geq (N + 1)$. Note that there are two trade-offs behind this equation:

a. The trade-off between the local tooth number and readout tooth location. If there are more local teeth, the location (frequency) of the readout teeth can be more flexible since there are more complete groups formed. Conversely, the requirement of the readout tooth location will be stricter if there are fewer local teeth. In practice, it is generally much more difficult to accurately control the frequency of the readout teeth with the precision of the repetition rate than to obtain more local/readout teeth. Therefore, the general practical solution could be to make the aggregate bandwidth of local and readout FC moderately larger than that of the target FC and to roughly control the frequency of the readout comb (e.g., with the precision of 0.1 nm). This is what we do in the experiment.
b. The trade-off between the number of teeth of the local FC and readout FC. As the equation suggests, fewer readout teeth are needed if there are more local teeth, and vice versa. It should be noted that, although in theory only the sum of the bandwidth of local FC and readout FC is regulated to realize the one-to-one mapping of the target teeth, a relatively broad local FC (short local pulse) will be more beneficial in practice, as it can provides a better time gating (Section 1.2 and 4.2) and a higher upconversion efficiency (Section 4.1).

*2.7 Bandwidth requirements for repetition rates and carrier–envelope offset frequency (CEO) frequencies*

In the last subsection, we discuss the bandwidth requirements on optical side. In this subsection, we discuss instead the requirements on RF side, specifically, $f_{r,L}, f_{r,T}, \delta, f_{ceo,L}$ and $f_{ceo,T}$. Without loss of generality, we continue the assumption that $f_{ceo,L} = 0$; thus, $f_{ceo,T}$ is effectively the relative $f_{ceo}$ between the target FC and local FC.

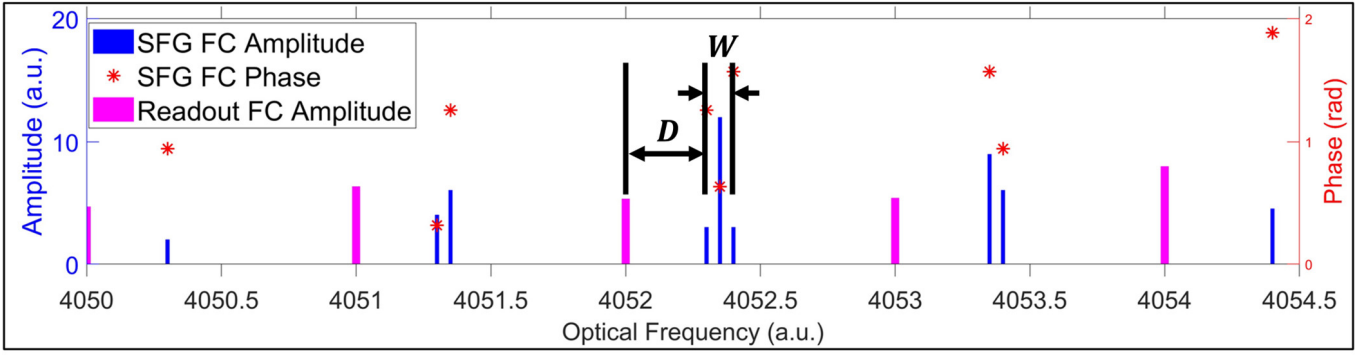
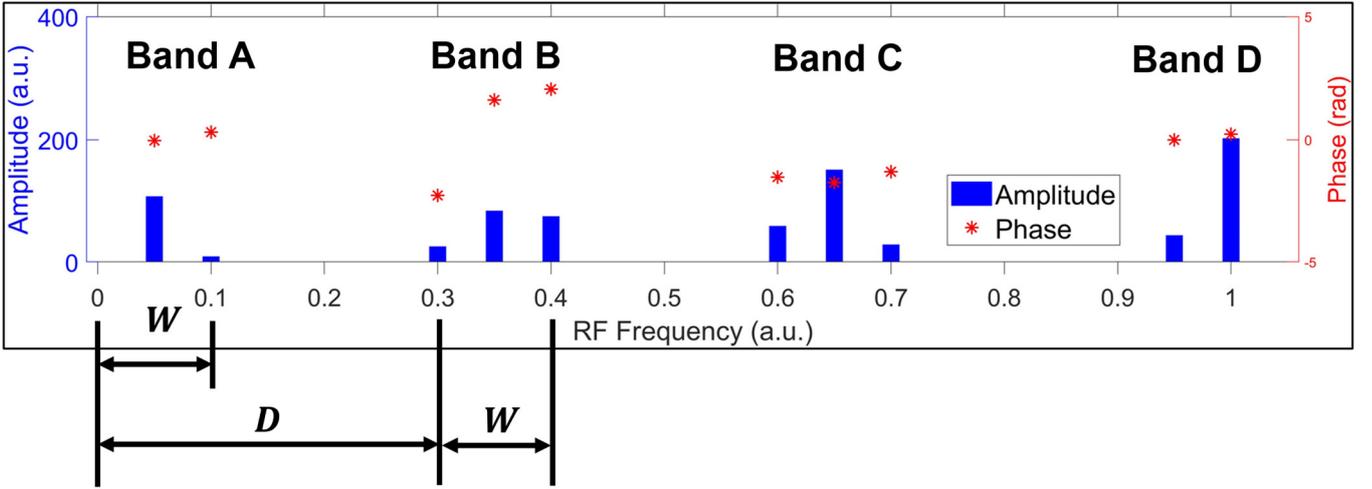

**Supplementary Fig. 2-3| Bandwidth requirements for RF frequencies.** The SFG FC and readout FC are from the Supplementary Fig. 2-1. D: the spectral distance (RF frequency) from the first tooth of an SFG group to its primary readout tooth. W: the spectral width of one complete group.

To quantify the requirements, here we define two important parameters (see the illustration in Supplementary Fig. 2-3):

a. The spectral (frequency) distance from the first tooth of an SFG group to its primary readout tooth, denoted by D. Note that the "first tooth of an SFG group" refers to the SFG tooth that corresponds to the first target tooth (the tooth with minimum frequency in the target FC).

$$D = mod\big((n_{first}f_{r,T} + f_{ceo,T}), f_{r,L}\big)$$

$mod(A, B)$ denotes the remainder after division of dividend A by divisor B, and $(n_{first}f_{r,T} + f_{ceo,T})$ is the optical frequency of the first tooth of the target FC.

b. The spectral width of one complete group, denoted by W.

$$W = (n_{last} - n_{first})\delta = BW_T \times \frac{\delta}{f_{r,T}}$$

$BW_T$ denotes the optical bandwidth of target FC.

Additionally, to realize a one-to-one mapping, two kinds of spectral overlap need to be avoided:

a. Avoiding overlap between band A(D) and band B(C), which requires:

$$D > W$$

b. Avoiding overlap between band B and band C, which requires:

$$D + W < \frac{f_{r,L}}{2}$$

Similar to dual-comb spectroscopy (DCS), $\frac{\delta}{f_{r,T}}$ needs to be small enough to provide enough bandwidth in the RF domain, i.e., to satisfy the requirement b. In addition, $f_{ceo,T}$ also need to be determined carefully to satisfy requirement a, which is different with DCS.

Note that the above bandwidth requirements are effective when a single detector is used for heterodyne photodetection. For the case that an ideal balanced detector is used, the requirements are simplified to only one equation:

$$W < \frac{f_{r,L}}{2}$$

This is because the band A and band D are eliminated by the balanced detector since they are common-mode signal from the SFG FC. In another word, the balanced detector can double the bandwidth for RF band B (C) assuming unchanged $\frac{\delta}{f_{r,T}}$, which makes the RF bandwidth requirement effectively same as the general dual-comb.

## 3. Comparison between different techniques: principle

In this section, we will compare DCS, C.W. upconversion spectroscopy, electric-optic sampling (EOS), and cross-comb spectroscopy (CCS) (Supplementary Fig. 3-1) using simple mathematical descriptions. Then, we will demonstrate that C.W. upconversion and EOS are essentially two special cases of the cross-comb; the former uses a very narrow-band local "FC" with only one "comb tooth", and the latter uses a very broadband local FC (very short local pulse) which also functions as the readout FC. We will describe them in both the time domain and the frequency domain. Especially, we show that the CCS in a general configuration can utilize the optical bandwidth in a more efficient way, compared to EOS. In all these techniques, if the full electric field profile of the readout FC (local FC) is available, generally acquired by field-resolved measurements (e.g., FROG), the electric field of the target FC can also be reconstructed based on measured correlation signal. This extra information could be helpful in some ways; however, it is not necessary for the goal of general absorption spectroscopy.

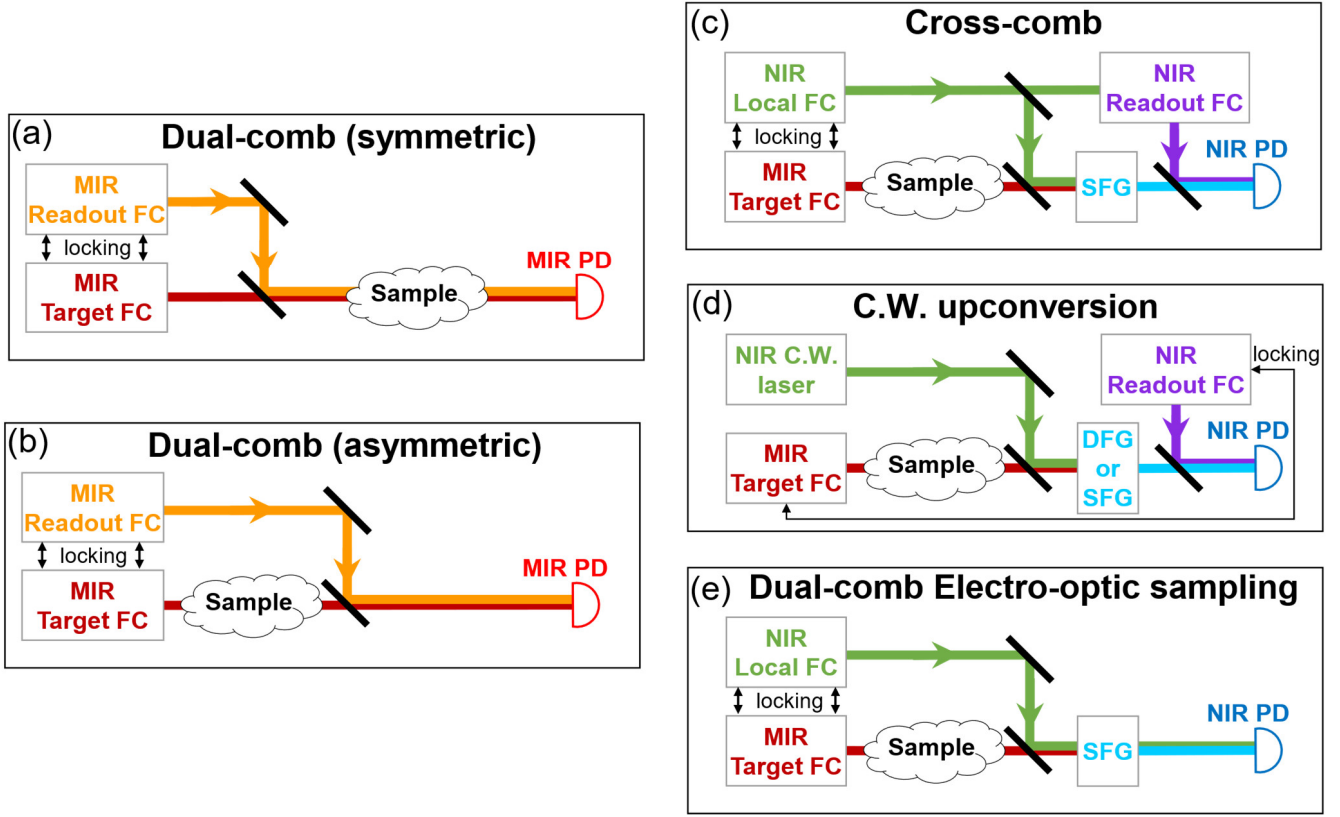

**Supplementary Fig. 3-1| Simplified schematics of different techniques.** Note that generally balanced detectors are used, which are simplified to be single detectors in the schematics. Also, there may be additional equipment before the detector, which is also omitted here; for example, an ellipsometry setup for electro-optic sampling (e).

To begin with, let us review the cross-correlation theorem:

$$C(\tau) = f(t) \otimes h(t) = \int_{-\infty}^{+\infty} f^*(t)h(t+\tau)dt \quad \Rightarrow \quad \mathcal{F}\{C(\tau)\} = F^*(\omega)H(\omega);$$

Or equally:

$$C(\tau) = f(t) \otimes h(t) = \int_{-\infty}^{+\infty} f(t)h^*(t-\tau)dt \quad \Rightarrow \quad \mathcal{F}\{C(\tau)\} = F(\omega)H^*(\omega)$$

Where $F(\omega)$ and $H(\omega)$ denote the Fourier transform of $f(t)$ and $h(t)$, respectively.

*3.1 DCS*

Firstly, for DCS with a symmetric (collinear) configuration (Supplementary Fig. 3-1(a))[8],

$$c(\tau) = \int_{-\infty}^{+\infty} e_T(t)e_R^*(t-\tau)dt$$

$$C(\omega) = \mathcal{F}\{c(\tau)\} = E_T(\omega)E_R^*(\omega)$$

where $e_T(t)$ and $e_R(t)$ denote the electric field of the target FC (pulse) and readout FC without passing the sample (passing the reference cell), while $E_T(\omega)$ and $E_L(\omega)$ denote their Fourier transform, respectively. $c(\tau)$ denotes the cross-correlation signal measured by the detector in the time domain, and $C(\omega)$ is its Fourier transform in the frequency domain.

Let assume the sample's spectral response is $S(\omega)$, including both spectral amplitude $|S(\omega)|$ and spectral phase $Ang\ (S(\omega))$. If we use $e(t)$ and $e'(t)$ to denote the electric field of a pulse before and after passing the sample, we have:

$$\mathcal{F}\{e'(t)\} = S(\omega)\mathcal{F}\{e(t)\} = S(\omega)E(\omega)$$

Therefore, for the cross-correlation signal $c'(\tau)$, measured when the target pulse and readout pulse pass the sample:

$$c'(\tau) = \int_{-\infty}^{+\infty} {e'}_T(t){e'}_R^*(t-\tau)dt$$

$$C'(\omega) = \mathcal{F}\{c'(\tau)\} = E_T(\omega)S(\omega)E_R^*(\omega)S'(\omega) = E_T(\omega)E_R^*(\omega)|S(\omega)|^2$$

By comparing those two measurements (with and without sample), we have:

$$D(\omega) = \frac{C'(\omega)}{C(\omega)} = |S(\omega)|^2$$

$D(\omega)$ denotes the comparison between those two measurements. It shows that this measurement can only provide spectral intensity of the sample's response, which lacks the phase information.

In fact, a symmetric DCS measurement is essentially a traditional FTIR (Michelson interferometer), which gives information only about spectral intensity but not spectral phase. Therefore, one cannot get any temporal information on the target pulses which are disturbed by the sample. In other words, the correlation signal $c(\tau)$ is independent of the spectral phase of ${e'}_T(t)$, which is cancelled as the readout pulse also passes the sample.

Secondly, for DCS with an asymmetric (dispersive) configuration (Supplementary Fig. 3-1(b)),

$$c(\tau) = \int_{-\infty}^{+\infty} e_T(t)e_R^*(t-\tau)dt$$

$$C(\omega) = \mathcal{F}\{c(\tau)\} = E_T(\omega)E_R^*(\omega)$$

$$c'(\tau) = \int_{-\infty}^{+\infty} e'_T(t)e_R^*(t-\tau)dt$$

$$C'(\omega) = \mathcal{F}\{c'(\tau)\} = E_T(\omega)S(\omega)E_R^*(\omega)$$

$$D(\omega) = \frac{C'(\omega)}{C(\omega)} = S(\omega)$$

Note that in this configuration, the readout pulse does not pass the sample before being combined with the target pulse. In this case, the measured $D(\omega)$ is dependent on the phase of $S(\omega)$; thus, one can get phase information of the sample response.

However, one still cannot recover the full electric field of the target pulse, $e_T(t)$ (or $e'_T(t)$) only by measurement of $C(\omega)$ (or $C'(\omega)$), in which $E_T(\omega)$ (or $E'_T(\omega)$) is modulated by $E_R^*(\omega)$. This is because $E_R^*(\omega)$ is generally unknown unless some other field-resolved measurements (e.g., FROG) are applied to measure it. Nonetheless, general absorption spectroscopy does not require the full knowledge of $e_T(t)$, since what we need to measure is $S(\omega)$ rather than $E_T(\omega)$, assuming $e_R(t)$ does not change for measurements with and without the sample.

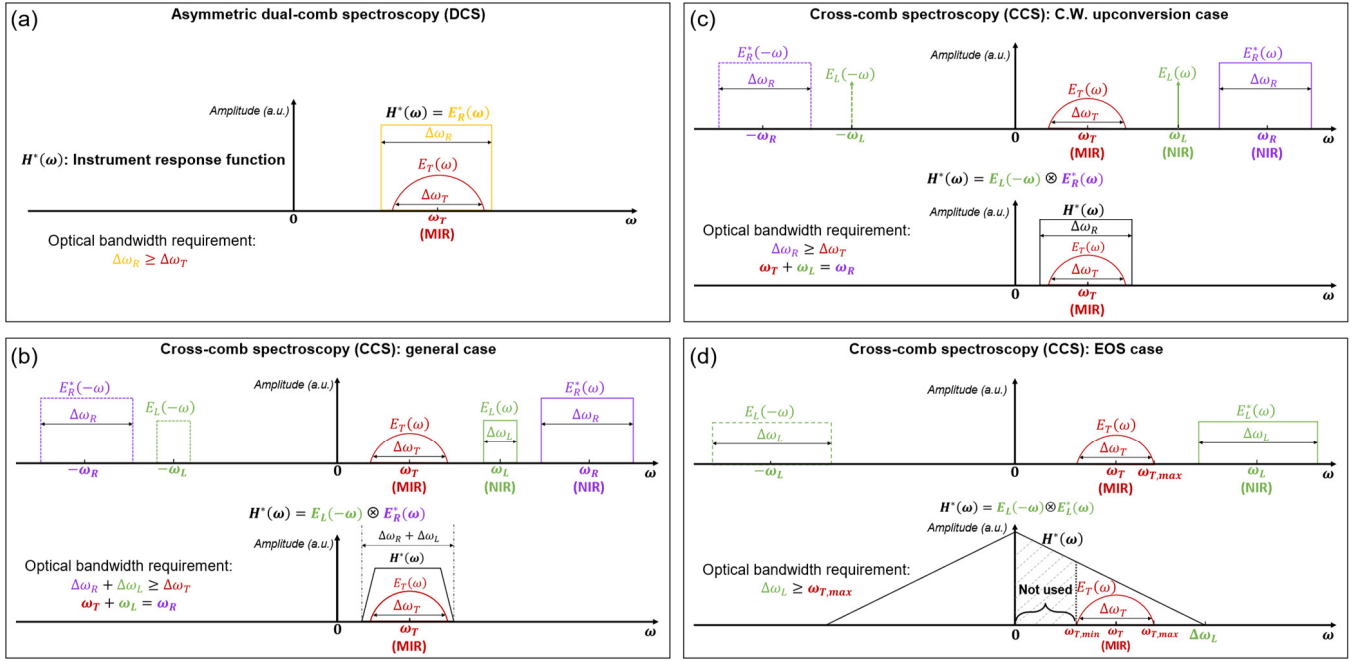

**Supplementary Fig. 3-2| Instrument response function of different techniques.** $\omega_T, \omega_L,$ and $\omega_R$: spectral center of the target FC, local FC, and readout FC. $\Delta\omega_T, \Delta\omega_L,$ and $\Delta\omega_R$: optical bandwidth of target FC, local FC, and readout FC. Note that we only plot the spectral amplitude (e.g., $|E_L(\omega)|$) for each function and assume zero spectral phase for all of them. Also, the spectral profiles of the local and readout FC ($E_L(\omega), E_R(\omega)$) are simplified to rectangular functions for clarity. **a.** asymmetric DCS. **b.** general CCS. The bandwidth requirement agrees with the result of our derivation in Section 2 in which comb teeth are included. **c.** CCS (C.W. upconversion case). **d.** CCS (EOS case). $\omega_{T,max}$ ($\omega_{T,min}$), the maximum (minimum) frequency of the target spectrum.

Supplementary Figure 3-2(a) illustrates the spectral amplitudes of $E_T(\omega)$ and $E_R^*(\omega)$ as well as the optical bandwidth requirement for the readout FC. $H^*(\omega)$ denotes the response function of the instrument, which is simply equal to $E_R^*(\omega)$ in this case. Note that this illustration, as well as the following illustrations for other techniques, depicts only the spectral envelopes and thus does not account for individual comb lines. This description simplifies the math without losing any generality since the comb lines can be understood effectively as sampling the envelopes in frequency domain (or equivalently, as a periodic extension in the time domain). The full description of CCS which factors in comb lines is presented in Section 2.

### 3.2 CCS
Thirdly, let us discuss CCS, which has the additional step of frequency conversion (Supplementary Fig. 3-1(c)).
Step 1: nonlinear upconversion
$$e_{SFG}(t,\tau) = e_T(t)e_L(t-\tau)$$
where $e_L(t)$ denotes the electric field of the local FC (pulse). Note that the phase-matching effect and nonlinear efficiency are not included here for simplicity.
Step 2: linear readout (same as asymmetric DCS)
$$c(\tau) = \int_{-\infty}^{+\infty} e_{SFG}(t,\tau)e_R^*(t-\tau)dt = \int_{-\infty}^{+\infty} e_T(t)e_L(t-\tau)e_R^*(t-\tau)dt$$

Let $h(t) = e_L^*(t)\,e_R(t)$, we can rewrite the above equation as:

$$c(\tau) = \int_{-\infty}^{+\infty} e_T(t)h^*(t-\tau)dt$$

$$C(\omega) = \mathcal{F}\{c(\tau)\} = E_T(\omega)H^*(\omega)$$

Above is the result for the measurement without sample, and for the measurement with sample we have:

$$c'(\tau) = \int_{-\infty}^{+\infty} e'_T(t)h^*(t-\tau)dt$$

$$C'(\omega) = \mathcal{F}\{c'(\tau)\} = E_T(\omega)S(\omega)H^*(\omega)$$

$$D(\omega) = \frac{C'(\omega)}{C(\omega)} = S(\omega)$$

Like asymmetric DCS, one can get phase information of the sample response, but $e_T(t)$ cannot be fully recovered since $E_T(\omega)$ is modulated by $H^*(\omega)$ in $C(\omega)$. However, this does not impede the measurement of the absorption spectrum $S(\omega)$.

In this case, the response function of the instrument is $H^*(\omega)$, based on that $h(t) = e_L^*(t)\,e_R(t)$, we have:

$$H^*(\omega) = E_L(-\omega) \otimes E_R^*(\omega)$$

which is illustrated in Supplementary Fig. 3-2(b).

### 3.3 C.W. upconversion and EOS

Both C.W. upconversion and EOS can be shown to be special cases of the above CCS description. To describe C.W. upconversion (Supplementary Fig. 3-1(d)), nothing needs to be modified in the CCS equations, except that $e_L(t-\tau)$ denotes a continuous sinusoidal wave instead of a pulse. Also, it should be noted that, using an SFG or DFG process for nonlinear upconversion does not make a fundamental difference here; the equations are equivalent up to a complex conjugation. The illustration is shown in Supplementary Fig. 3-2(c).

EOS (Supplementary Fig. 7(e)) requires a more careful discussion. Let us start with equations of CCS.

Step 1: nonlinear upconversion

$$e_{SFG}(t,\tau) = e_T(t)e_L(t-\tau)$$

In the case of ideal EOS, $e_L(t)$ is much shorter than $e_T(t)$. In other words, in the temporal span of $e_L(t)$, $e_T(t)$ varies very little and can be approximated to be constant. Thus, we have:

$$e_{SFG}(t,\tau) = \boldsymbol{e_T(t)}e_L(t-\tau) \cong \boldsymbol{e_T(\tau)}e_L(t-\tau)$$

Another way to interpret this is that $e_L(t)$ is approximated to be a Dirac delta function ($\delta(t-\tau)$) that samples $e_T(t)$ in the time domain.

With this approximation, we can continue to derive the next readout step. Note that in EOS the role of readout pulse is played by the local pulse itself.

Step 2: linear readout

$$c(\tau) = \int_{-\infty}^{+\infty} \boldsymbol{e_T(t)}e_L(t-\tau)e_R^*(t-\tau)dt \cong \int_{-\infty}^{+\infty} \boldsymbol{e_T(\tau)}e_L(t-\tau)e_L^*(t-\tau)dt = \boldsymbol{e_T(\tau)}\int_{-\infty}^{+\infty} e_L(t-\tau)e_L^*(t-\tau)dt = K\boldsymbol{e_T(\tau)}$$

$$C(\omega) = \mathcal{F}\{c(\tau)\} = KE_T(\omega)$$

$$C'(\omega) = \mathcal{F}\{c'(\tau)\} = KE_T(\omega)S(\omega)$$

$$D(\omega) = \frac{C'(\omega)}{C(\omega)} = S(\omega)$$

where $K$ denotes the constant that equals to the integration $\int_{-\infty}^{+\infty} e_L(t-\tau)e_L^*(t-\tau)dt$, the core of which is independent of the parameter delay $\tau$. As shown in the equation, under this approximation, the correlation signal $c(\tau)$ is equal to the electric field of target pulse $e_T(\tau)$ up to a constant. Thus, under the approximation of the ideal local pulse (infinitely short pulse width), one can obtain the full electric field of the target pulse $e_T(t)$ in addition to the absorption spectrum $S(\omega)$.

In practice, the finite pulse duration of the sampling pulse always imposes a frequency-dependent instrument response[9,10], which is illustrated in Supplementary Fig. 3-2(d). In this case, the instrument response function $H^*(\omega)$ is the "autoconvolution" of the local spectrum.

$$H^*(\omega) = E_L(-\omega) \otimes E_L^*(\omega)$$

In contrast to DCS and CCS, EOS needs the bandwidth of the local FC ($\Delta\omega_L$) to be equal or larger than the maximum frequency of the target FC ($\omega_{T,max}$) to detect the full spectrum of the target FC. This explains why EOS requires a much broader optical bandwidth compared to DCS and CCS. However, the $H^*(\omega)$ band below the minimum frequency of the target FC ($\omega_{T,min}$), is not effectively utilized, resulting from the fact that the same continuous FC is used as both the readout and local FC.

### 3.4 C.W. upconversion and EOS described by comb-teeth mapping

In the previous subsection, we have described C.W. upconversion spectroscopy and dual-comb EOS using the language of CCS without including comb teeth. In this subsection, we do the same thing factoring in comb teeth, following the derivation in Section 2.5. Note that Fig. 4 of the main paper is a good illustration for this subsection.

Based on what we derived for RF band B in Section 2.5, we can write the general formula for $j_{th}$ target tooth mapped in RF band B:

$$I_j = \left(\sum_{m=1}^{M} L_m R_{m+j}^*\right) X_j$$

where M denotes the total number of local teeth. Note that all the subscripts denote effective tooth index.

For the case of C.W. upconversion, there is only one "local tooth", so the formula is simplified to be

$$I_j = L_1 R_{m+1}^* X_j$$

Everything can be described well by the language of CCS.

For the case of ideal EOS, let us review the approximation that we made in the time domain, which is:

"In the span of $e_L(t)$ or $e_R(t)$ (very short local/readout pulse), $e_T(t)$ (target pulse) varies slowly, and thus can be approximated as constant."

Correspondingly, in the frequency domain, we can have such an equivalent approximation:

"In the span of $E_T(\omega)$ (very narrowband, relatively), $E_L(\omega)$ or $E_R(\omega)$ (very broadband, relatively) varies slowly and can be approximated as constant."

With this approximation, we have:

$$R_m \cong R_{m+1} \cong R_{m+2} \cong R_{m+3} \cdots \cdots \cong R_{m+N}$$

where N denote the total number of target teeth. Thus, we have:

$$\text{for any } j, \sum_{m=1}^{M} L_m R_{m+j}^* \cong K$$

where K denotes a constant.

$$I_j = \left(\sum_{m=1}^{M} L_m R_{m+j}^*\right) X_j \cong K X_j, \text{for any } j$$

This result is equivalent to the equation $C(\omega) = \mathcal{F}\{c(\tau)\} = KE_T(\omega)$, which we derived in the last subsection in the time domain. Both results show that, in the limit of ideal EOS, the measured correlation signal (RF heterodyne beating) is equal to the electric field of the target pulse up to a constant.

The case of nonideal EOS is well demonstrated in reference [5].

In summary, both C.W. upconversion spectroscopy and EOS fall into the category of CCS, representing two opposite limits on the bandwidth of the local comb.

## 4. Comparison between different techniques: performance

In the last section, we compared how different techniques work. With the same model and assumptions, in this section, we further compare some of their important metrics, including detection bandwidth, efficiency, SNR and dynamic range. We will first compare detection bandwidth and efficiency of different upconversion methods, pointing out that the short-pulse CCS is overall more efficient. Secondly, we will present a comparison between DCS and short-pulse CCS in terms of SNR and dynamic range, highlighting the effect of temporal gating[3]. Lastly, some insights into the design rules of CCS systems are provided, based on the results of this section.

### *4.1 Detection bandwidth and efficiency of upconversion methods*

#### 4.1.1 General CCS, symmetric CCS, and C.W. upconversion CCS

In last section, we defined a response function $H(\omega)$ to describe different methods. Here, we will continue to use it for a more quantitative comparison. With the same mathematical assumptions as before (Supplementary Fig. 3-2), the dimensions of $H(\omega)$ of the general CCS case are calculated, based on the given parameters (heights and width) of $E_L(\omega)$ and $E_R(\omega)$ (Supplementary Fig. 4-1(a)).

In practice, it is more general and convenient to discuss and compare spectral intensity (power) instead of spectral amplitude of the electric field. Thus, the amplitude spectrum is squared to the intensity spectrum, and a power gain function $G(\omega) = |H^*(\omega)|^2$ is defined to describe the detection efficiency of the system (Supplementary Fig. 4-1(b)).

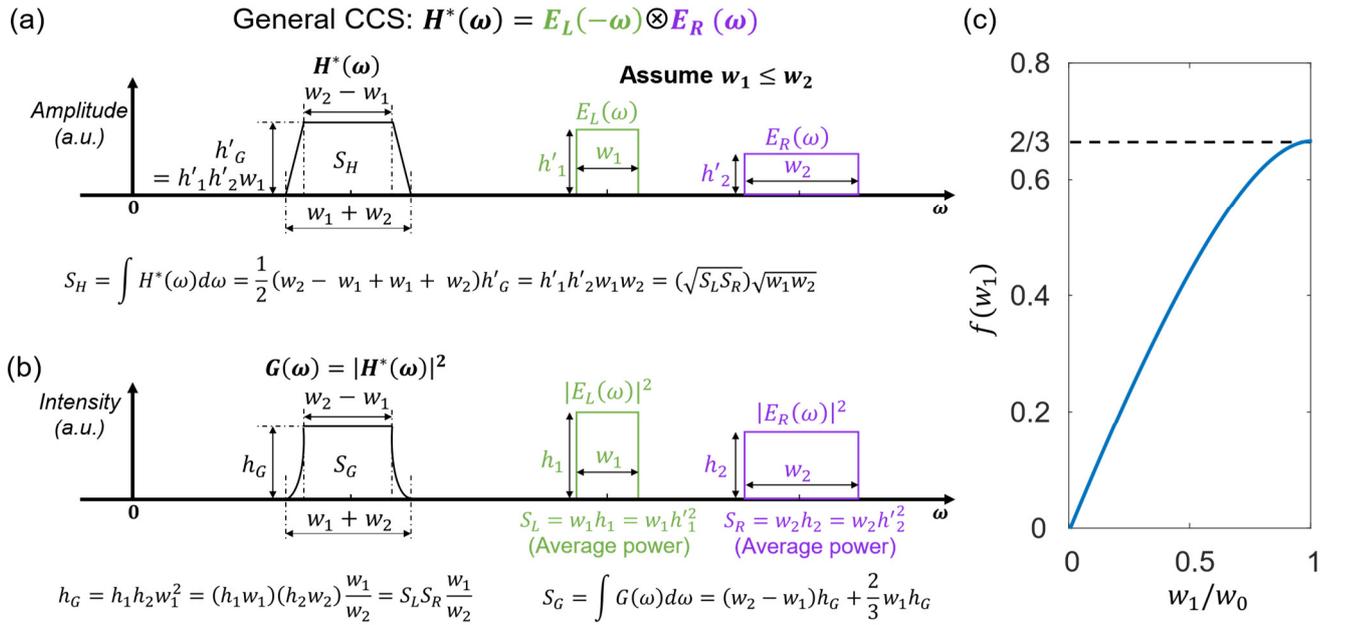

**Supplementary Fig. 4-1| Instrument response function and gain function of general CCS. a.** Instrument response function $H(\omega)$ spectral amplitude of $E_L(\omega)$ and $E_R(\omega)$. **b.** Detection gain function $G(\omega)$ and spectral intensity of $E_L(\omega)$ and $E_R(\omega)$. $S_{L(R)} = \int |E_{L(R)}(\omega)|^2 d\omega$. $w$ and $h$ denote the width and height of those spectral profiles, respectively. We assume $w_1$ (bandwidth of local FC) $\leq w_2$ (bandwidth of readout FC) in this derivation, with no lack of generality for conclusions we draw. **c.** $f(w_1) = \left(2w_0 - \frac{4}{3}w_1\right)\frac{w_1}{2w_0 - w_1}$. It monotonically increases with $w_1$ on the interval $[0, w_0]$.

Three metrics of $G(\omega)$ are used to quantify it:

$h_G$: the maximum value ("height") of $G(\omega)$, which describe the highest detection gain at the center part of $G(\omega)$.

$w_G$: the bandwidth of $G(\omega)$. Here we simply use zero points to define the width.

$S_G$: the area under $G(\omega)$, i.e., $\int G(\omega) d\omega$. This quantifies the total gain of the system.

Under our assumption of rectangular spectral profiles for $E_L(\omega)$ and $E_R(\omega)$, we get:

$$h_G = h_1 h_2 w_1^2 = (h_1 w_1)(h_2 w_2)\frac{w_1}{w_2} = S_L S_R \frac{w_1}{w_2}$$

$$BG = w_1 + w_2$$

$$S_G = \int G(\omega) d\omega = (w_2 - w_1) h_G + \frac{2}{3} w_1 h_G = \left(w_2 - \frac{1}{3} w_1\right) h_G = \left(w_2 - \frac{1}{3} w_1\right) \frac{w_1}{w_2} S_L S_R$$

where we assume $w_1 \leq w_2$.

Please refer to Supplementary Fig. 4-1 and its caption for a detailed definition of variables. Note that $S_{L(R)}$, which denotes the area under $|E_{L(R)}(\omega)|^2$, is equivalent to the average power of the local (readout) FC.

There are in total four effective free variables: $S_L, S_R, w_1, and\ w_2$. For reasonable comparison later, let us assume a fixed total bandwidth $w = 2w_0 = w_1 + w_2$ and a fixed total power $S = 2S_0 = S_L + S_R$ for the local and readout FC.

Let us first consider the choice of $S_L$ and $S_R$. Based on the equation of $h_G$, we have

$$S_L S_R \leq \left(\frac{S_L + S_R}{2}\right)^2 = S_0^2, where\ "="\ holds\ when\ S_L = S_R = S_0.$$

Thus, we want to make $S_L = S_R = S_0$ to optimize $h_G$ (this also optimizes $S_G$). In this case, $S_L S_R = S_0^2$, which is a constant. Secondly, let us consider the choice of $w_1$ and $w_2$. To optimize $h_G$, it is obvious that we want to make $w_1 = w_2 = w_0$, thus we have $h_G = S_L S_R = S_0^2$.

As for $S_G$, a more careful calculation is needed. Since $S_L S_R$ is already set to a constant, let us consider function $f(w_1, w_2) = \left(w_2 - \frac{1}{3} w_1\right) \frac{w_1}{w_2}$ with the constraint that $w_1 + w_2 = w = 2w_0$, by which $f(w_1, w_2)$ is effectively a function of one variable ($w_1$ or $w_2$). Making it a function of only $w_1$, we plot $f(w_1)$ in Supplementary Fig. 4-1(c). $f(w_1)$ monotonically increases with $w_1$ on the interval $[0, w_0]$, and reaches its maxima of 2/3 at $w_1 = w_2 = w_0$. Based on these observations, we name the case where $S_L = S_R = S_0$ and $w_1 = w_2 = w_0$ as "symmetric CCS."

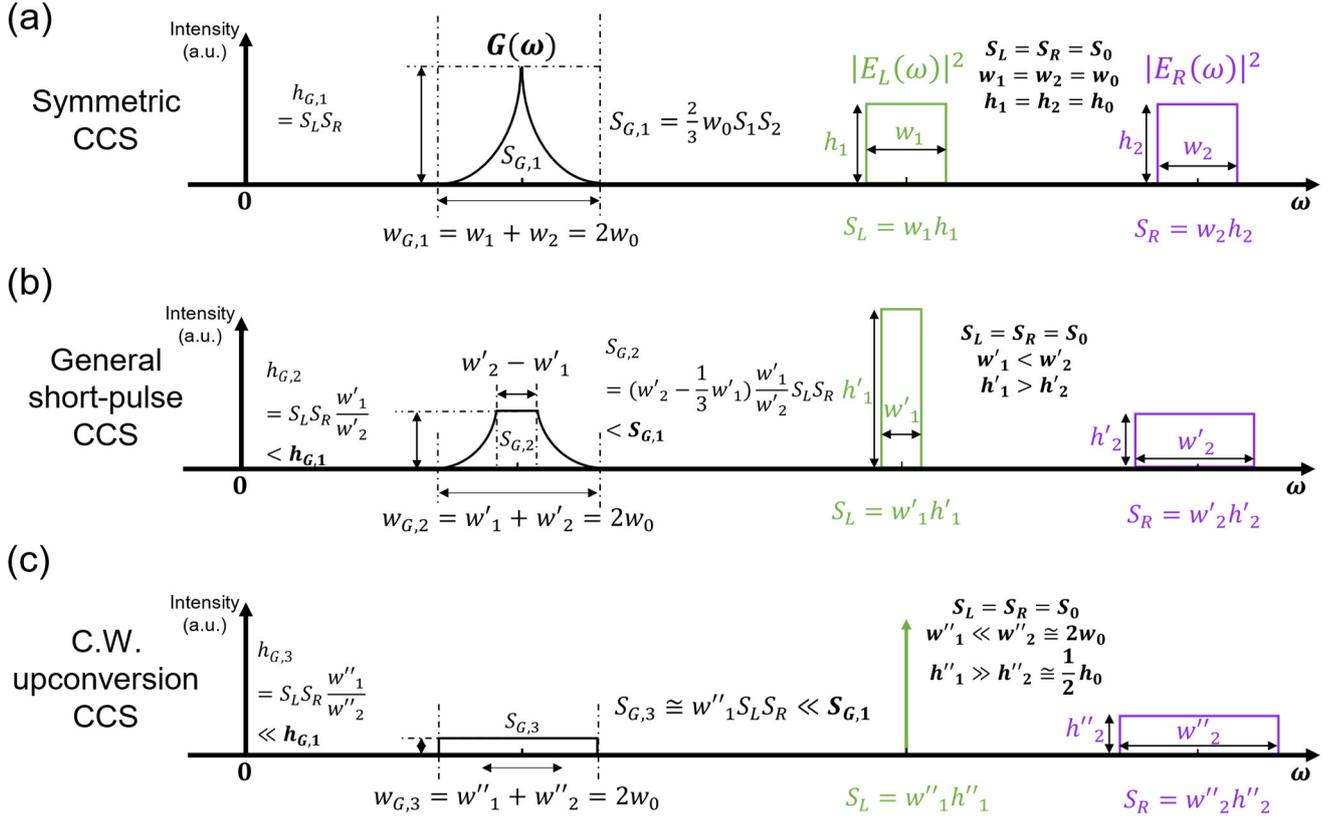

**Supplementary Fig. 4-2| Gain function $G(\omega)$ of symmetric CCS (a), general CCS (b), and C.W. upconversion CCS (c).** In (c), the local FC spectral intensity profile is like a Dirac-delta function, which has a very small width $w''_1$ and a very large height $h''_1$, while their product $S_L$ is kept same as the other two cases.

With the derivation above, it is easy to compare the symmetric CCS, general short-pulse CCS (with parameters close to symmetric CCS) and C.W. upconversion CCS, shown in Supplementary Figure. 4-2 (a)-(c), respectively. Note that we keep the total bandwidth and power of local and readout FC the same for all three cases. As the bandwidth of the local FC shrinks, symmetric CCS becomes general short-pulse CCS, which finally becomes C.W. upconversion CCS. During the transition, although the bandwidth $w_G$ stays the same, the maximum gain $h_G$ and the total area $S_G$ monotonically decrease. For the C.W. upconversion case (panel c), the bandwidth of the local FC shrinks to $w''_1$ that is $\ll w_0$. If we compare $h_G$ and $S_G$ between symmetric case and C.W. upconversion case, we have:

$$\frac{h_{G,optim}}{h_{G,C.W.}} = \frac{h_{G,1}}{h_{G,3}} = \frac{2w_0}{w''_1} \gg 1$$

$$\frac{S_{G,optim}}{S_{G,C.W.}} = \frac{S_{G,1}}{S_{G,3}} = \frac{2w_0}{3w''_1} \gg 1$$

In short, general short-pulse CCS (where local and readout FC both have a broad bandwidth) has a much higher detection gain than C.W. upconversion CCS. This should not be a surprising result if we think about the comparison of conversion efficiency of general nonlinear optics processes between C.W. laser and short pulses. The enhancement ratio here is exactly equivalent to the peak power enhancement ratio between coherent short pulses and a C.W. laser with the same average power. The conclusion we arrive at here has its roots in the same reason why people prefer short pulses over C.W. lasers for nonlinear optics: the much higher peak power of short pulses.

### 4.1.2 Symmetric CCS and EOS CCS

Here we compare short-pulse CCS with EOS CCS, the former of which is represented by the symmetric CCS, and the comparison is illustrated in Supplementary Figure. 4-3. Still, we keep their total bandwidth and power the same to make a fair comparison.

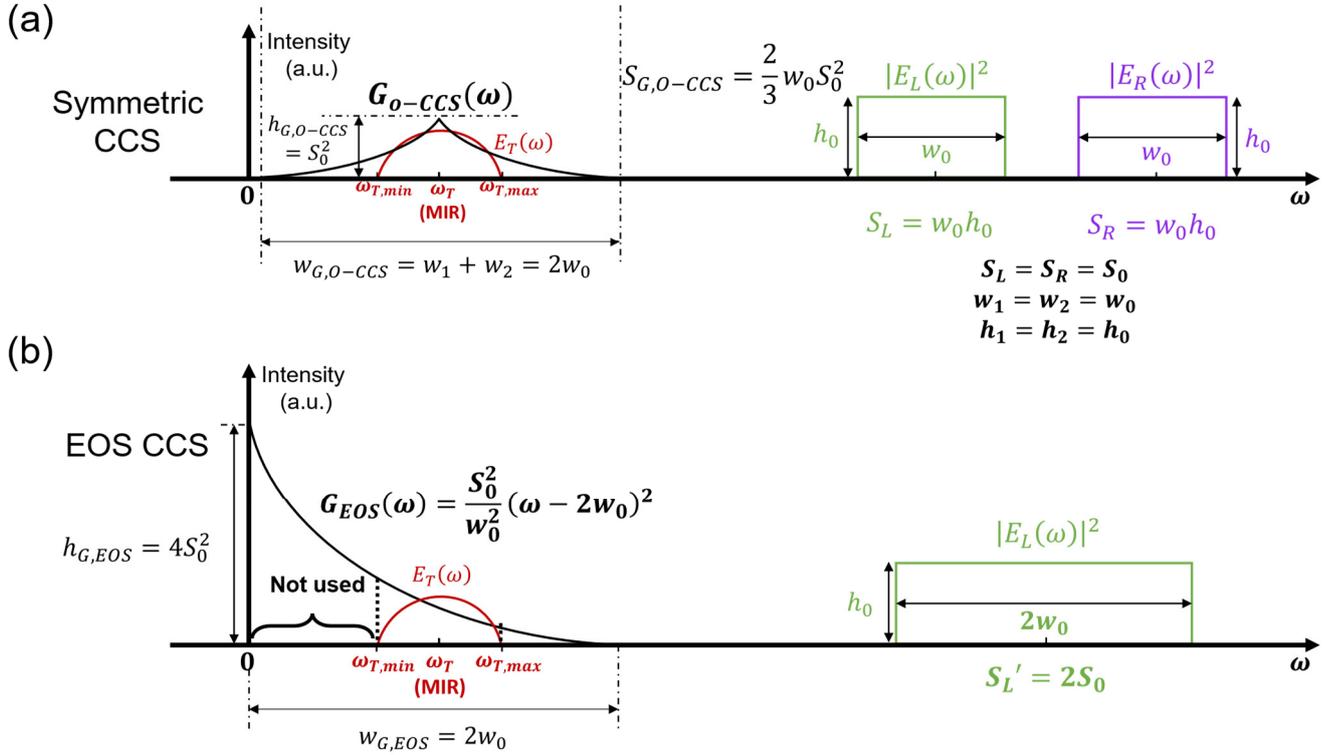

**Supplementary Fig. 4-3| Gain function $G(\omega)$ of symmetric CCS (a) and EOS CCS (b).**

Based on the calculation shown in the plot, although the maxima of $G_{EOS}(\omega)$, i.e., $h_{G,EOS}$, is larger than that of the $G_{ccs}(\omega)$ i.e., $h_{G,O-CCS}$, the $h_{G,EOS}$ is at $\omega = 0$, which cannot overlap with target spectrum at all. Actually, $G_{EOS}(\omega)$ monotonically decreases with $\omega$, and the largest part of it in amplitude, i.e., $\omega \in [0, \omega_{T,min}]$, is not effectively used, as shown in the Supplementary Fig. 3-2. A system response curve with a similar shape and trend has been calculated in ref [9] (see Fig. S1 in its Supplementary Materials). This qualitative behavior already shows that EOS uses resources in a less efficient way compared to general CCS. Nevertheless, we still proceed to give a more quantitative description.

For a fair comparison, we want to calculate $G_{EOS}(\omega_{T,min})$ (the gain at $\omega = \omega_{T,min}$), $G_{EOS}(\omega_T)$ (the gain at the center of the target spectrum, $\omega = \omega_T$), and $S_{G,EOS} = \int_{\omega_{T,min}}^{\omega_{T,max}} G_{EOS}(\omega) d\omega$ (the overall gain that covers target spectrum). Based on the model presented before, we can have an analytical expression for $G_{EOS}(\omega)$:

$$G_{EOS}(\omega) = \frac{S_0^2}{w_0^2}(\omega - 2w_0)^2$$

Of course, the value of the three metrics depends on the value of $\omega_{T,min}$, $\omega_T$, and $\omega_{T,max}$, and different values can result in different conclusions for the comparison. Here, to give an example, let us adopt some values that are close to our experiments.

Let us set $\omega_T = 75\ THz\ (4\ \mu m)$ and assume we are going to detect a 30-THz-broad target bandwidth, that is, $\omega_{T,min} = 60\ THz\ (5\ \mu m)$ and $\omega_{T,max} = 90\ THz\ (3.33\ \mu m)$. If we set the center of the local FC at $200\ THz\ (1.5\ \mu m)$, the bandwidth of the local FC needs to be at least 90 THz, i.e., $w_0 = 45\ THz$, in order to detect the whole target spectrum. This suggests the local FC must span $155\ THz - 245\ THz\ (1.22\ \mu m\ to\ 1.93\ \mu m)$, and is therefore nontrivial to generate. Meanwhile, symmetric CCS only requires 15-THz local and readout FCs (in total 30 THz), which are much less challenging experimentally.

With the assumed values of $w_0, \omega_{T,min}, \omega_T$, and $\omega_{T,max}$, we have $\omega_{T,min} = \frac{4}{3}w_0, \omega_T = \frac{5}{3}w_0, and\ \omega_{T,max} = 2w_0$. Then the three metrics can be calculated and compared.

$$G_{EOS}(\omega_{T,min}) = \frac{4}{9}S_0^2; \quad G_{EOS}(\omega_T) = \frac{1}{9}S_0^2; \quad G_{EOS}(\omega_{T,max}) = 0; \quad S_{G,EOS} = \int_{\omega_{T,min}}^{\omega_{T,max}} G_{EOS}(\omega)d\omega = \frac{8}{81}w_0 S_0^2$$

$$\frac{G_{EOS}(\omega_{T,min})}{G_{o-CCS}(\omega_T)} = \frac{4}{9}; \quad \frac{G_{EOS}(\omega_T)}{G_{o-CCS}(\omega_T)} = \frac{1}{9}; \quad \frac{S_{G,EOS}}{S_{G,o-CCS}} = \frac{4}{27} \cong 0.15$$

In this case, the maximum value of $G_{EOS}(\omega)$ is just 4/9 of that of the $G_{o-CCS}(\omega)$, and this value is only at the left edge of the target spectrum. If we instead compare the gain at the center of the two functions, the ratio becomes only 1/9. Since they are of different profiles, it is more reasonable to compare their overall gain $S_G$, and $S_{G,EOS}$ is only ~15% of $S_{G,o-CCS}$. In short, although EOS may require much more experimental effort, its overall detection efficiency can be much lower than symmetric (general short-pulse) EOS.

4.1.3 Summary

In this subsection we compare symmetric CCS, general short-pulse CCS, C.W. upconversion CCS, and EOS CCS. Compared to C.W. upconversion CCS, general short-pulse CCS can have a much higher detection (upconversion) efficiency, which comes from the enhancement of peak power of short pulses over C.W. laser. Compared to EOS CCS, general short-pulse CCS (represented by symmetric CCS) can have a much larger detection bandwidth and detection efficiency, although it is much less experimentally demanding. Overall, among different upconversion configurations, short-pulse CCS has advantages in bandwidth, efficiency, flexibility, and experimental complexity.

*4.2 Background comparison between DCS and CCS: sensitivity, SNR, and dynamic range*

4.2.1 Overview

In this part, we will provide a quick qualitative description. In asymmetric DCS, we have the correlation signal:

$$c(\tau) = \int_{-\infty}^{+\infty} e_T(t)e_R^*(t-\tau)dt$$

Note that only the cross term, i.e., the effective correlation signal, is kept in this equation. The background that is omitted in the equation is:

$$B = \int_{-\infty}^{+\infty} |e_T(t)|^2 dt + \int_{-\infty}^{+\infty} |e_R^*(t-\tau)|^2 dt = \int_{-\infty}^{+\infty} |e_T(t)|^2 dt + \int_{-\infty}^{+\infty} |e_R^*(t)|^2 dt$$

This background is equal to the sum of the full power of the target pulse and local pulse, which is independent of the delay, $\tau$. At large delay $\tau$, when the weak tail (optical free induction decay) of the target pulse is being sampled by the local pulse, the effective correlation signal can be much smaller than the constant background. In other words, the extra noise incurred by the background from the strong target pulse can envelop the weak useful signal at the tail. Even in the absence of technical noise, the strong background can saturate the detector, thus fundamentally limiting the dynamic range and SNR of the measurement[3].

In CCS, in which a short local pulse is used (not necessarily as short as in the EOS case), the correlation signal is:

$$c(\tau) = \int_{-\infty}^{+\infty} e_{SFG}(t,\tau)e_R^*(t-\tau)dt = \int_{-\infty}^{+\infty} e_T(t)e_L(t-\tau)e_R^*(t-\tau)dt$$

The omitted background terms are:

$$B(\tau) = \int_{-\infty}^{+\infty} |e_{SFG}(t,\tau)|^2 dt + \int_{-\infty}^{+\infty} |e_R^*(t-\tau)|^2 dt = \int_{-\infty}^{+\infty} |e_T(t)\boldsymbol{e_L(t-\tau)}|^2 dt + \int_{-\infty}^{+\infty} |e_R^*(t)|^2 dt$$

In stark contrast to DCS, the background in CCS is dependent on the delay $\tau$, as the target pulse is "temporally gated" by a short local pulse $e_L(t-\tau)$. At the weak tail of the target pulse, where the effective correlation signal is weak, the background is also very weak, as it is free from the strong power of the center (peak) part of the target pulse. This allows a much stronger target pulse to be used, which promises a higher SNR at the weak tail, compared to the linear DCS. This behavior is well shown qualitatively in Fig. 1(e) of the main paper.

It is readily seen that the temporal gating effect is better as the local pulse is shorter. Also, a shorter local pulse benefits the upconversion efficiency. This is one of the reasons why we use a relatively short local pulse (broadband local FC) in our experiment, although only the total bandwidth of the local FC and readout FC is regulated in theory to fully map the target FC.

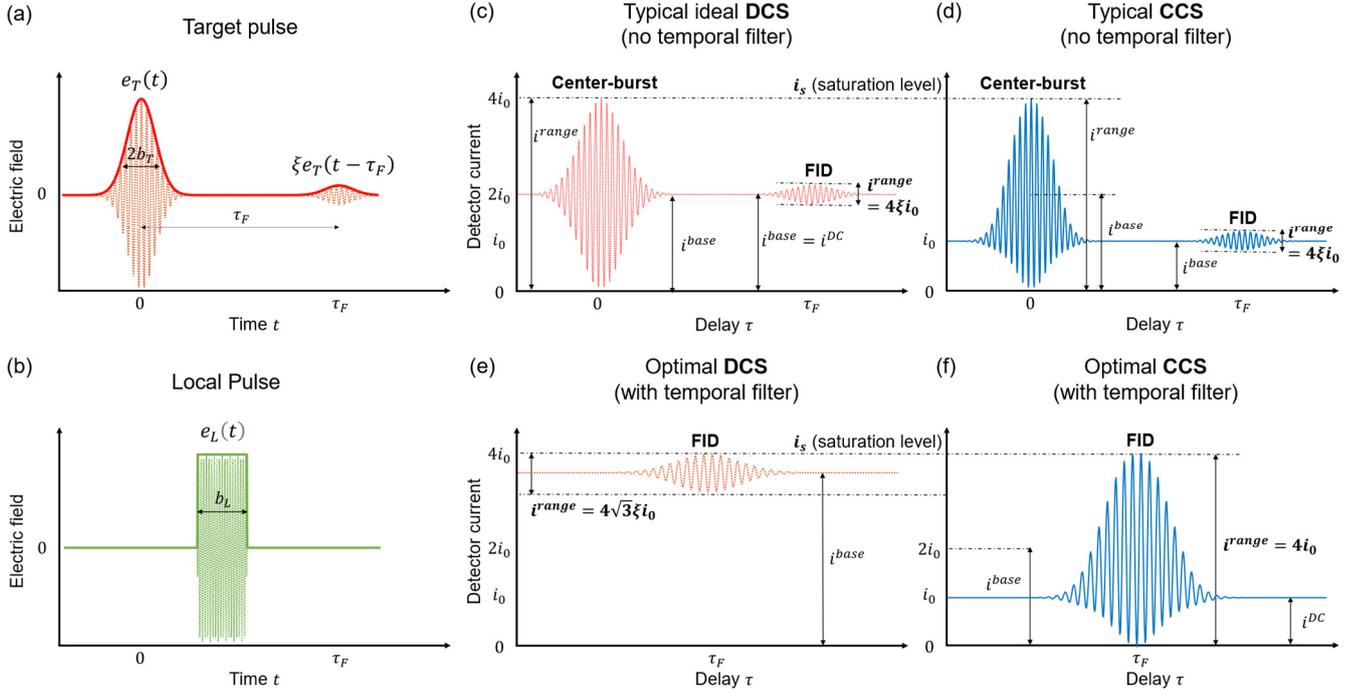

**Supplementary Fig. 4-4| Comparison between CCS and DCS. a.** Electric field of the target pulse $e_T(t)$, followed by its FID part $\xi e_T(t-\tau_F)$. **b.** Electric field of the local pulse $e_L(t)$. **c-d.** Typical interferograms of DCS **(c)** and CCS **(d)**, without temporal filter. **e-f.** Optimal interferograms of DCS **(e)** and CCS **(f)** at FID, with a temporal filter to remove their center-bursts.

### 4.2.2 Assumptions of the model

To compare the SNR and sensitivity of DCS and short-pulse CCS quantitatively, firstly, we need to set up our model (Supplementary Fig. 4-4):

(1) When the target pulse does not go through sample, we describe it by: $e_T(t) = a g_b(t) \exp(i\omega_T t)$, where $a_0$ denotes the amplitude, $g_b(t)$ denotes the pulse envelope function. $g_b(t)$, and other functions $g(t)$ to follow, is set to be a gaussian function with a width of $b$, i.e., $\exp\left(-\left(\frac{t}{b}\right)^2\right)$. Also, we assume a slowly varying envelope, i.e., $\frac{dg(t)}{dt} \ll \omega$.

(2) After the target pulse goes through the sample, we describe it by $e_T^s(t) = \left[a_T g_{b_T}(t) + \xi a_T g_{b_T}(t-\tau_F)\right]\exp(i\omega_T t) = e_T(t) + \xi e_T(t-\tau_F)$ (see panel (a)). The first term denotes the original probing pulse (center-burst), and the second term denotes the FID from the sample. $\xi$ denotes the amplitude ratio between the center and FID, which is $\ll 1$ if assuming a weak absorption measurement. $\tau_F$ denotes the time interval between the FID and the pulse center, and we assume it is $\gg b_T$ (the pulse width of $a_{\tau_T}(t)$), i.e., the FID signal is far enough from the center thus the field amplitude here is not influenced by the term $g_{b_T}(t)$. Indeed, it is not physically sound to assume the FID signal has the same pulse shape as the original excitation pulse. However, what matters for the following calculations is the amplitude ratio ξ between the center-burst and the FID signal, and these assumptions simplify the math without changing the core of the calculation. According to the derivation in part II of the supplementary material of ref.[3], $\xi$ approximately equals the absorption up to a few other factors.

(3) For DCS, we set the readout pulse similar to the target: $e_R(t) = a_R g_{b_R}(t) \exp(i\omega_T t)$ (see $e_T(t)$ in panel (a)). For simplicity we assume $b_R = b_T$.

(4) For CCS, we set the local pulse as a square gating function, i.e., $e_L(t) = rect_{b_L}(t)$. $b_L$ denotes the width of this square function, for which we assume $\tau_F \gg b_L > b_T$ (i.e., a local pulse width larger than target pulse width but much smaller than the interval between the FID and center-burst). Therefore, for the target pulse $e_T^s(t)$, around the center-burst, we have $e_{SFG}(t, \tau \sim 0) = e_T^s(t)e_L(t-\tau) \cong e_T(t)$; around the FID, we have $e_{SFG}(t, \tau \sim \tau_F) = e_T^s(t)e_L(t-\tau) \cong \xi e_T(t-\tau_F)$. This assumption means the gating function effectively separates different temporal parts of the target pulse, and only the part that overlaps with the gating function can influence the detector value at a specific delay $\tau$. As for the readout pulse, we assume it has the same envelope as the readout of DCS, although with a different optical frequency $\omega$.

(5) For both DCS and CCS, we use a single slow detector that samples at a rate $f_s$, the same as the repetition rate of the readout FC ($f_s = f_{r,R} = f_{r,L} \cong f_{r,T}, T_s = \frac{1}{f_s} \ggg \tau_F \gg b$). Taking DCS as an example, at a delay $\tau$, the detector current can be represented by: $i(\tau) = C_{ip} \int_{-\frac{T_s}{2}}^{\frac{T_s}{2}} |e_T(t) + e_R(t-\tau)|^2 dt$, where $C_{ip}$ denotes a constant parameter that converts the result of the integration (equivalent to optical power) to photocurrent. Note that $C_i$ includes some physical constants related to the electric field as well as the quantum efficiency and responsivity of the detector, which are not main subjects of this study. The parameter, $C_{ip}$, and integration limits, $\frac{T_s}{2}$ and $-\frac{T_s}{2}$, will be omitted to simplify the equation later, since the calculation is not generally sensitive to them.

(6) There are three kinds of noise that would be generally included in the SNR discussion: detector noise (NEP), shot noise, and relative intensity noise (RIN)[11]. For clarity, we do not include the RIN in this calculation. Therefore, unless the optical power is very low, shot noise is the main noise source here. Since we are going to apply the idea of "temporal gating", we study the SNR of the measurement in the time domain.

Let us first start with a typical ideal DCS measurement (no FID). We assume $e_T(t) = e_R(t) = ag_b(t)\exp(i\omega_T t)$, and $i_a = \int e_T^2(t)dt$.

When $|\tau| \gg b$, i.e., the two pulses do not overlap, and we have a background signal:
$$i^{DC} = \int e_T^2(t)dt + \int e_R^2(t)dt = 2i_a.$$

At $\tau = 0$, when the two pulses constructive interfere, i.e., the maxima of the interferogram, we have:
$$i^+(0) = \int (2e_T(t))^2 dt = 4i_a.$$

At $\tau \cong 0$, when the two pules destructively interfere, i.e., the minimum of the interferogram, we have:
$$i^-(0) \cong \int [e_T(t) - e_R(t)]^2 dt = 0$$

Thus, the range of the interference here, denoted by $i^{range}$, is $4i_a$, which can be understand as the amplitude of the "signal". To evaluate the noise, we define the base current $i^{base}$ around $\tau = 0$ as the average value of $i^+(0)$ and $i^-(0)$, which is:
$$i^{base}(0) = \frac{1}{2}(i^+(0) + i^-(0)) = 2i_a.$$

In this case, the base current around the maxima is equal to the background $i_{DC}$. Since shot noise here increases with $i_b$, to optimize SNR, one wants to optimize the ratio $i^{range}/i^{base}$. Actually, this ratio is equivalent to the "interferometric visibility" up to a factor of $1/2$.

The shot noise around $\tau = 0$ can be expressed by:

$$i^{sn} = C_{isn}\sqrt{i^{base}} = C_{isn}\sqrt{2i_a}$$

$C_{isn}$ denote a constant parameter that convert $\sqrt{i^{base}}$ into current noise. Like $C_{ip}$, $C_{isn}$ includes some physical constants which are not the main subjects of this study.

If the optical power is low, the dominant noise is detector noise, and the SNR of the measurement is:

$$SNR = \frac{i^{range}(0)}{i_{NEP}} = \frac{4i_a}{i_{NEP}}$$

The dominant noise becomes the shot noise when the optical power is higher, and SNR of the measurement is:

$$SNR = \frac{i^{range}(0)}{i^{sn}} = \frac{4i_a}{C_{isn}\sqrt{2i_a}}$$

Apparently, the SNR increases with $\sqrt{i_a}$, and this dependence agrees with general shot-noise-limited measurement since $i_a \propto$ the power of the target (or readout) pulse. Also, the above two equations are equivalent to equation (24) and (25) of ref [11], although different letter conventions are used, and the RIN and dynamic range terms are not included.

The SNR can be improved by increasing $i_a$ (average power of FCs), which stops when $i^+(0) = 4i_a$ reaches $i_s$, i.e., the saturation level of the detector. Let $i_0 = \frac{1}{4}i_s$, thus the SNR reaches maxima when $i_a = i_0$ ($4i_0 = i_s$):

$$SNR_{DCS,max} = \frac{4i_0}{C_{isn}\sqrt{2i_0}}$$

### 4.2.3 SNR at FID, without temporal filter

To compare the sensitivity for weak absorption, we need to calculate SNR at the FID instead of the center-burst. Let us start with DCS following the last subsection. This time we assume $e_T^s(t) = [a_0 g_b(t) + \xi a_0 g_b(t - \tau_F)]\exp(i\omega_T t) = e_T(t) + \xi e_T(t - \tau_F)$, $e_R(t) = e_T(t) = a_0 g_b(t)\exp(i\omega_T t)$, $i_0 = \int e_R^2(t)dt = \int e_T^2(t)dt$.

At the center-burst, $i^+(0) = 4i_0 = i_s$, and the SNR here reaches maxima, as explained above.

However, at the FID, everything changes. We have:

$$i^+(\tau_F) = \int ((1+\xi)e_T(t))^2 dt + \int e_T^2(t)dt = [(1 + 2\xi + \xi^2) + 1]i_0$$

$$i^-(\tau_F) = \int ((1-\xi)e_T(t))^2 dt + \int e_T^2(t)dt = [(1 - 2\xi + \xi^2) + 1]i_0$$

$$i^{range}(\tau_F) = i^+(\tau_F) - i^-(\tau_F) = 4\xi i_0$$

$$i^{base} = \frac{1}{2}[i^+(\tau_F) + i^-(\tau_F)] = (2 + \xi^2)i_0 \cong 2i_0$$

$$SNR_{DCS,FID} = \frac{4\xi i_0}{C_{isn}\sqrt{2i_0}} = \xi SNR_{DCS,max}$$

In short, although the amplitude of the signal (the interference) is $\xi$ times weaker than that at the center-burst, the noise level is still the same since the detector can see all the background from the center-burst, i.e., the large energy which contributes to only the noise here but not the signal.

Let us then consider CCS. At the center-burst, since we assume $e_{SFG}(t, \tau \sim 0) = e'_T(t)e_L(t - \tau) \cong e_T(t)$, its SNR has the same results as DCS. However, at the FID, $e_{SFG} \cong \xi e_T(t - \tau_F)$, we have:

$$i^+(\tau_F) = \int \left((1 + \xi)e_R(t)\right)^2 dt = [(1 + 2\xi + \xi^2)]i_0; \quad i^-(\tau_F) = \int \left((1 - \xi)e_R(t)\right)^2 dt = [(1 - 2\xi + \xi^2)]i_0$$

$$i^{range}(\tau_F) = 4\xi i_0; \quad i^{base} \cong i_0$$

$$SNR_{CCS,FID} = \frac{4\xi i_0}{C_{isn}\sqrt{i_0}} = \sqrt{2}\, SNR_{DCS,FID}$$

This shows a $\sqrt{2}$ SNR enhancement of CCS over DCS, which is not significant. In fact, here we greatly limit the capability of CCS. On one hand, in this simple model we assume the same detector performance (NEP, responsibility, quantum efficiency, saturation level, and etc.) of the NIR detector of CCS and MIR detector of DCS, the latter of which should be worse the former. On the other hand, we still limit the optical power, especially the local pulse, of the CCS, to avoid detector saturation at the center-burst, which is not necessary for detecting weak absorption.

### 4.2.4 SNR at FID, with temporal filter

Now we introduce a temporal filter that throws out the interference signal at the center-burst (before the FID) and only keep the interference around FID for the detection of weak absorption[3]. In this case, we no longer care about the detector saturation at the center-burst; the limits of the SNR at the FID are set by the detector saturation at the FID locally.

For DCS, we can increase both the power of $e_R(t)$ and $e_T(t)$ by 2 times, i.e., $e_R(t) = \sqrt{2}e_T(t), e_T^S(t) = \sqrt{2}e_T(t) + \sqrt{2}\xi e_T(t - \tau_F)$, to maximize the SNR here. We cannot increase more since $i^+(\tau_F)$ already saturates the detector. We have:

$$i^+(\tau_F) = \int \left((\sqrt{2} + \sqrt{2}\xi)e_T(t)\right)^2 dt + \int 2e_T^2(t)dt = [(2 + 4\xi + 2\xi^2) + 2]i_0 \cong 4i_0 = i_s$$

$$i^-(\tau_F) = \int \left((\sqrt{2} - \sqrt{2}\xi)e_T(t)\right)^2 dt + \int 2e_T^2(t)dt = [(2 - 4\xi + 2\xi^2) + 2]i_0 \cong 4i_0 = i_s$$

$$i^{range}(\tau_F) = i^+(\tau_F) - i^-(\tau_F) = 8\xi i_0; \quad i^{base}(\tau_F) = \frac{1}{2}[i^+(\tau_F) + i^-(\tau_F)] = (4 + 2\xi^2)i_0 \cong 4i_0$$

$$SNR_{DCS,FID}^{gated} = \frac{8\xi i_0}{C_{isn}\sqrt{4i_0}} = \sqrt{2}\, SNR_{DCS,FID} = \sqrt{2}\xi SNR_{DCS,max}$$

Compared to not-gated case, the SNR can only be increased by a factor of $\sqrt{2}$.

For CCS, thanks to temporal gating by the local pulse, the detector signal at the FID is free from the power of the center part of the target pulse (the second term of the RHS of the above equation of $i^+(\tau_F)$, i.e., $\int 2e_T^2(t)dt$). Thus, we can increase the power of the local pulse by a factor of $(\frac{1}{\xi})^2$, i.e., $e'_L(t) = \frac{1}{\xi}rect_{b_L}(t)$. Then, at the FID,

$$e_{SFG}(\tau = \tau_F) = \xi e_T(t - \tau_F)e'_L(t - \tau_F) = \xi e_T(t - \tau_F)\frac{1}{\xi}rect_{b_L}(t - \tau_F) \cong e_T(t - \tau_F)$$

If we keep $e_R(t) = e_T(t)$, we will have an SNR equivalent to that of the center-burst of a typical ideal DCS measurement, calculated in section 4.2.2:

$$i^+(\tau_F) = \int \left(2e_T(t - \tau_F)\right)^2 dt = 4i_0 = i_s; \quad i^-(\tau_F) = 0; \quad i^{range}(\tau_F) = 4i_0; \quad i^{base}(\tau_F) = 2i_0$$

$$SNR_{CCS,FID}^{gated} = \frac{4i_0}{C_{isn}\sqrt{2i_0}} = \frac{1}{\sqrt{2}\xi}SNR_{DCS,FID}^{gated}$$

Therefore, a $\frac{1}{\sqrt{2}\xi}$ SNR enhancement is demonstrated for CCS compared to DCS at FID. The weaker the absorption, the stronger the enhancement can be. The reason for this lies in the fact that the DCS signal at the FID always comes with a factor of $1/\xi$ stronger DC background which contributes only to the noise but not the signal, while CCS does not.

In practice, one can "infinitely" decreases $\xi$ by decreasing the sample concentration or gas cell length. However, the enhancement ratio $\frac{1}{\sqrt{2}\xi}$ cannot be infinitely increased; the limit is set by two factors, whichever comes first:

1. The SFG efficiency. To fully reach the SNR enhancement, one needs to upconvert the target pulse by a factor $1/\xi$ stronger using local pulses with a higher peak power. However, this can be clamped by the highest available peak power of local pulses or the damage threshold of the SFG crystal.
2. The damage threshold of the NIR detector. Although the detector saturation at the center-burst no longer matters if we discard the signal there, we still do not want the power there to damage the detector. Generally, however, the detector is damaged by the average power rather than the peak power, and the average power on the detector mainly depends on that of the readout pulses instead of local pulses. In other words, just a very small portion of the local pulse average power contributes to the total average power on the detector, the ratio of which is decided by the "duty cycle" of the SFG process (see Section 1.3). In temporal-filtered CCS, the strategy is to use stronger local pulses while keeping the readout pulses unchanged, which adds only a tiny optical average power on the detector. Hence, it is very unlikely that the limit of this factor comes earlier than the former.

4.2.5 Sensitivity and dynamic range, with temporal filter

Following the SNR calculation, a comparison of sensitivity (the minimal detectable $\xi$) becomes straightforward. Let us define $\xi_{min}$, the minimal detectable $\xi$ that makes SNR=1. For DCS at the FID, we have:

$$SNR_{DCS,FID}^{gated} = \frac{8\xi_{min}^{DCS} i_0}{C_{isn}\sqrt{4i_0}} = 1$$

$$\xi_{min}^{DCS} = \frac{C_{isn}}{2\sqrt{i_s}}$$

Clearly it is limited by $i_s$, the detector saturation. ($i_s = 4i_0$)

For CCS, the sensitivity depends entirely on the upconversion capability, as discussed before. Let us assume the upconversion conversion ratio is $1/\xi_0$. Then for an absorption $\xi_0$, as derived before, we have an SNR $= \frac{4i_0}{C_{isn}\sqrt{2i_0}}$ at the FID if we set $e_R(t) = e_T(t)$. However, this SNR is much larger than 1 and more than enough to detect the absorption signal. Thus, we can further decrease the absorption. Let us assume an extra absorption factor $\xi_1$ to make the target FID a "small signal". Simultaneously, to maximize SNR, we set $e_R(t) \cong 4\, e_T(t)$. Then, at the FID, we have:

$$i^{range}(\tau_F) = 16\xi_1 i_0;\ i^{base}(\tau_F) = 4i_0;$$

$$SNR_{CCS,FID} = \frac{16\xi_1 i_0}{C_{isn}\sqrt{4i_0}}$$

Assuming this SNR=1, we have:

$$\xi_{1,min} = \frac{C_{isn}}{4\sqrt{i_s}} = \frac{1}{2}\xi_{min}^{DCS}$$

Adding the ratio $\xi_0$ back, we have:

$$\xi_{min}^{CCS} = \xi_0 \xi_{1,min} = \frac{1}{2}\xi_0 \xi_{min}^{DCS}$$

Therefore, CCS can detect a $\frac{1}{2}\xi_0$ smaller absorption compared to DCS. The detection is limited by the upconversion capability rather than the detector.

It should be noted that, in order to maximize the sensitivity (minimize the detectable $\xi$) for both techniques, we set their parameters ($e_R(t)$ and $e_T(t)$) to make $i_{base}$ close to detector saturation. However, in such settings, there will not be dynamic range for the detection: a larger absorption will excess the saturation limit, and lower absorption will make $SNR < 1$ thus not detectable. In short, for a detection where we want to get good dynamic range, we do not want to use the settings for the best sensitivity.

It can be much more complicated to give a complete description of their dynamic range. Hence, here we give a qualitative comparison based on our discussion before. Let assume we set the $e_R(t)$ for both techniques small, such that the noise that the beating signal needs to overcome on the lowest absorption limit is the detector noise instead of the shot noise. For CCS, as the absorption increases, the detectable beating signal can vary from $i_{NEP}$ to $i_s$. In other words, the dynamic range of the CCS is basically equal to that of the detector, i.e., $DR_{CCS} = DR_{detector} = i_s/i_{NEP}$. In DCS, on the lowest absorption limit, to make $\xi_{low}$ detectable, a factor of $1/\xi_{low}$ extra background needs to be put on the detector to make the beating signal overcome $i_{NEP}$. Thus, the range that the absorption $\xi$ can scale is $\xi_{low}i_s/i_{NEP}$, i.e., $DR_{DCS} = \xi_{low}i_s/i_{NEp} = \xi_{low}DR_{detector}$. In short, DCS has a factor of $\xi_{low}$ less dynamic range than CCS due to the extra background power. The lower the absorption (better sensitivity) you wish to detect with DCS, the larger the part of the detector dynamic range that must be sacrificed. However, this undesired tradeoff does not exist for CCS. A more comprehensive discussion will be the subject of our future work.

### 4.2.6 Summary

The discussion above demonstrates the fact that the sensitivity of short-pulse CCS is limited by upconversion capability (SFG efficiency), which is fundamentally different from DCS, where it is limited by the detector saturation. The beauty of the short-pulse-upconversion CCS is that strong local pulses can greatly enhance the peak power of the signal of interest in a localized temporal window, with minimal increase to the background signal and average power on the detector which add to noise and saturation of the detector. This time gating effect endows CCS advantages in SNR, sensitivity, and dynamic range.

Moreover, it should be noted that, among the three different upconversion configurations, only short-pulse-upconversion CCS can fully have these advantages. Firstly, C.W. upconversion CCS is basically DCS, which does not have advantages we discussed here at all. Secondly, for EOS CCS, one may expect it to have the same advantages since it also uses short pulses for upconversion, but this is nevertheless not true in terms of SNR and sensitivity. Admittedly, the even higher peak power of the local pulse (because of shorter pulse length) used in EOS CCS can provide even higher upconversion efficiency. However, when the average power of the local pulses is increased to detect weaker absorption, that of the readout part of the local spectrum is also increased, which can saturate the detector unexpectedly. In other words, in short-pulse-upconversion CCS, you can always use higher local power to amplify a weaker FID signal while keeping the readout power unchanged, and you will never saturate the detector. However, in EOS CCS, you cannot do the same since local and readout are from the same pulse (spectrum), and thus their power cannot be tuned independently.